\documentclass{emulateapj}
\usepackage{apjfonts}
\usepackage{graphicx}
\usepackage{epstopdf}
\newcommand{\secpoint}{\mbox{$''\mskip-7.6mu.\,$}}
\newcommand{\nar}{NewAR}

\slugcomment{Received 5/21/14; accepted 9/16/14}

\begin{document}

\title{A Spectroscopic Survey of WISE-selected Obscured Quasars with the Southern African Large Telescope}

\shorttitle{SALT Spectroscopy of WISE-Selected Obscured Quasars}
\shortauthors{HAINLINE ET AL.}

\author{\sc Kevin N. Hainline, Ryan C. Hickox, Christopher M. Carroll}
\affil{Department of Physics and Astronomy, Dartmouth College, Hanover, NH 03755}

\author{\sc Adam D. Myers, Michael A. DiPompeo}
\affil{Department of Physics and Astronomy, University of Wyoming, Laramie, WY 82071}

\author{\sc Laura Trouille}
\affil{Center for Interdisciplinary Exploration and Research in Astrophysics (CIERA) and Department of Physics and Astronomy, Northwestern University, Evanston, IL 60208, The Adler Planetarium, Chicago, IL 60605}

\author{}

\begin{abstract}

We present the results of an optical spectroscopic survey of a sample of 40 candidate obscured quasars identified on the basis of their mid-infrared emission detected by the Wide-Field Infrared Survey Explorer (WISE). Optical spectra for this survey were obtained using the Robert Stobie Spectrograph (RSS) on the Southern African Large Telescope (SALT). Our sample was selected with WISE colors characteristic of AGNs, as well as red optical to mid-IR colors indicating that the optical/UV AGN continuum is obscured by dust. We obtain secure redshifts for the majority of the objects that comprise our sample (35/40), and find that sources that are bright in the WISE W4 (22$\mu$m) band are typically at moderate redshift ($\langle z \rangle = 0.35$) while sources fainter in W4 are at higher redshifts ($\langle z \rangle = 0.73$). The majority of the sources have narrow emission lines, with optical colors and emission line ratios of our WISE-selected sources that are consistent with the locus of AGN on the rest-frame $g-z$ color vs. [NeIII]$\lambda$3869 / [OII]$\lambda\lambda$3726+3729 line ratio diagnostic diagram. We also use empirical AGN and galaxy templates to model the spectral energy distributions (SEDs) for the objects in our sample, and find that while there is significant variation in the observed SEDs for these objects, the majority require a strong AGN component. Finally, we use the results from our analysis of the optical spectra and the SEDs to compare our selection criteria to alternate criteria presented in the literature. These results verify the efficacy of selecting luminous obscured AGNs based on their WISE colors. 

\end{abstract}


\section{Introduction}
\label{sec:intro}

	A complete understanding of the evolution of galaxies requires a detailed exploration of the role played by active galactic nuclei (AGNs), highly luminous objects found in the centers of galaxies that are powered by accretion onto a  supermassive black hole. The most powerful AGNs, quasars, have been proposed to influence the interstellar gas content and temperature throughout a galaxy, which may lead to reduced star formation \citep[for recent reviews, see][]{fabian2012,alexander2012}. The line-of-sight orientation between the central supermassive black hole and any obscuring dust is thought to produce the array of different observed AGNs \citep{antonucci1993, urry1995}. This ``unification theory'' separates AGNs into two broad classes: obscured AGNs, where nuclear and galaxy-scale dust blocks emission from near the central black hole, and unobscured AGNs, where there exists a line of sight down to the central engine. It is far more straightforward to detect unobscured AGNs, while the full population of obscured AGNs is not fully characterized. Historically, unobscured AGNs have been selected by observing broad emission lines emitted from excited gas near the accreting supermassive black hole, and are often called Type I AGNs to differentiate them from Type II AGNs whose spectra only contain narrow emission lines. The difficulty in finding obscured AGNs is important given the evidence for an evolutionary link between obscured and unobscured AGNs. One of the primary proposed methods for fueling a quasar is through major mergers of gas-rich galaxies, where gas is funneled into the central regions of the galaxy \citep{kauffmann2000,hopkins2006,hopkins2008}. It is thought that quasar fueling can also lead to an obscuration of the UV and optical emission from the accretion process by the in-falling gas and dust, which would eventually be lifted by quasar radiative feedback, setting up a large-scale connection between the supermassive black hole and galaxy-wide star formation (Chen et al. in prep). Understanding the full population of obscured AGNs is vital for exploring both the cosmological growth of supermassive black holes as well as the connection to their host galaxies and larger-scale galaxy environments. 

	Deep X-ray surveys using both \emph{XMM-Newton} and \emph{Chandra} have revealed extensive samples of AGNs with a wide luminosity range out to large cosmic distances \citep{mainieri2002,alexander2003,mateos2005,brandt2010,xue2011}. AGN population synthesis models, however, require a substantial number of obscured AGNs to explain the shape of the cosmic X-ray background spectrum \citep{comastri1995, gilli2007, treister2009,ballantyne2011,shi2013}, and many of these objects have been missed in existing X-ray surveys \citep{worsley2005,tozzi2006, alexander2008, burlon2011}. A complimentary approach is to target obscured AGNs selected at mid-infrared (mid-IR) wavelengths, where obscuration effects are minimized and reprocessed light from dust near the central supermassive black hole produces a characteristic red IR power-law spectrum. \citet{lacy2004} and \citet{stern2005} described robust color criteria for separating AGNs from star-forming galaxies using the IR photometric bands targeted with the Infrared Array Camera (IRAC) instrument on the \emph{Spitzer Space Telescope}, which was further explored in \citet{donley2008,donley2012} and \citet{mendez2013}. \citet{hickox2007} studied a large sample of obscured and unobscured objects in the Bo{\"o}tes field selected to be AGNs by their IRAC colors. They demonstrated a bimodality in optical to mid-IR color that they used to separate obscured and unobscured sources.  \citet{lacy2013} presented an optical and near-IR spectroscopic study of a large sample of 786 objects selected with \emph{Spitzer} IRAC colors indicative of AGN activity and demonstrated the power in using \emph{Spitzer} selection techniques to target AGNs and obscured quasars even out to $z \sim 2 - 3$. However, IRAC-based studies like these are limited to only the relatively small fields observed by \emph{Spitzer}. Recently, the all-sky mid-IR photometric coverage from the \emph{Wide-field Infrared Survey Explorer} \citep[WISE;][]{wright2010} has also been used by many authors to separate active galaxies from star-forming galaxies. WISE covered the entire sky at four wavelengths: 3.4, 4.6, 12, and 22 $\mu$m (referred to as W1, W2, W3, and W4 throughout this work), reaching 5$\sigma$ point-source sensitivities better than 0.08, 0.11, 1, and 6 mJy in each photometric band, respectively. This unprecedented all-sky coverage makes WISE a powerful tool for selecting large numbers of obscured active galaxies for follow-up studies.
	
	There has been a wealth of research into the selection of AGNs by their WISE colors. \citet{assef2010}, \citet{ashby2009}, and \citet{stern2012} described WISE selection criteria that use only the W1$-$W2 color to separate AGNs from star-forming galaxies. \citet{stern2012} used WISE data from the COSMOS field to demonstrate that a WISE color cut of $\mathrm{W1}- \mathrm{W2} > 0.8$ is successful at recovering 78\% of \emph{Spitzer}-selected AGNs. A less conservative color cut, such as $\mathrm{W1}- \mathrm{W2}> 0.5$ suggested by \citet{ashby2009}, was shown to increase the completeness of the AGN sample, but the recovered sample suffered from more contamination from star-forming galaxies. \citet{mateos2012} used a mid-IR-selection method which relies on three of the WISE photometric bands:  W1, W2, W3. Using X-ray selected AGNs from the Bright Ultrahard \emph{XMM-Newton} survey, the authors demonstrated that the completeness of their MIR-selection method is strongly dependent on AGN luminosity, especially as compared to the luminosity of the host galaxy.
		
	While many of the methods used to define WISE obscured AGN selection rely on exploring the WISE colors of large \emph{existing} samples of active galaxies, it would be of interest to test these selection methods using a blind survey of obscured objects selected solely by their colors. By performing follow-up spectroscopy, we can test obscured AGN selection criteria, and explore both the type of objects and the redshift distribution of objects with red IR colors. Surveys of this nature are useful for understanding the full population of WISE-selected obscured quasars, as it is necessary to employ the large statistical sample of WISE quasars to study spatial clustering \citep[e.g.][DiPompeo et al. 2014]{donoso2013} as well as the role that AGNs play in galaxy evolution. Additionally, the all-sky coverage of WISE can be used to target rare objects of which there are few in existing spectroscopic fields.

	In this paper, we present long-slit optical spectroscopy obtained with the Southern African Large Telescope (SALT) for a sample of 40 objects selected with WISE and SDSS colors indicative of obscured quasar activity. We describe the full sample and our selection criteria in Section \ref{sec:sample}, examine the optical spectra and redshift distribution for the objects in Section \ref{sec:opticalspectra}, and in Section \ref{sec:linediagnostics} we discuss the use of various optical emission line diagnostics to examine the physical characteristics of objects that comprise the sample. We present the results of modeling the SEDs of the sample with a combination of AGN and star-forming galaxy templates in Section \ref{sec:sedmodeling}. We examine individual objects in our sample in Section \ref{sec:individualobjects}, compare the selection criteria we used to other recent WISE selection criteria in Section \ref{sec:agnselection}, and finally, we draw conclusions in Section \ref{sec:conclusions}. Throughout, we assume a standard $\Lambda$CDM cosmological model with $H_0 = 71$ km s$^{-1}$ Mpc$^{-1}$, $\Omega_{M} = 0.27$, and $\Omega_{\Lambda} = 0.73$ \citep{komatsu2011}.
	
\section{Quasar Sample, Observations and Data Reduction}
\label{sec:sample}

	\begin{figure*}[htbp]
	\epsscale{1.05} 
	\plotone{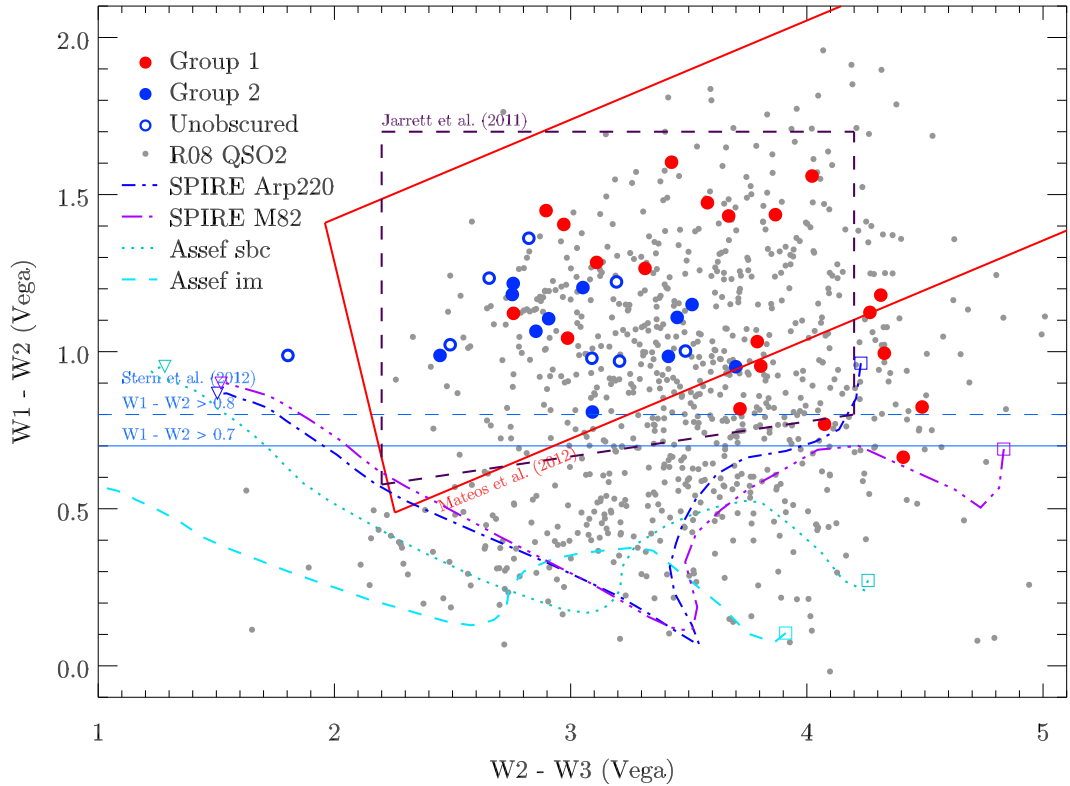}
	\caption{
	\label{fig:wisecolorcolor} Plot showing WISE W1$-$W2 color against W2$-$W3 color. The objects in our SALT sample (as measured with updated AllWISE photometry) are plotted with red (Group 1) and blue (Group 2) circles, and Type II quasars from \citet{reyes2008} are shown with small light grey circles. Open circles are used to plot objects where the SED modeling (outlined in Section \ref{sec:sedmodeling}) indicates that they are unobscured in the optical. We also plot various AGN selection criteria from the literature. We plot the AGN selection box described in \citet{jarrett2011} with a purple dashed line, we plot both the W1$-$W3 color cuts described in \citet{stern2012} with blue solid and dashed horizontal lines, and finally we plot the AGN-selection wedge from \citet{mateos2012} with a red solid line. While the bulk of our sample fall into the AGN selection regions, individual objects lie outside the different regions. We explore this in more detail in Section \ref{sec:agnselection}. To demonstrate the WISE colors of star-forming galaxies as a function of redshift, we plot tracks showing how the WISE colors for the \citet{assef2010} Sbc (cyan dotted line), Im (cyan dashed line), and \citet{polletta2007} Arp220 (blue dot-dashed line) and M82 (purple triple-dot dashed line) templates change as a function of redshift, where $z = 0$ is represented by an open square, and $z = 2$ is represented by an open downward-facing triangle. The most extreme star-forming galaxy M82 does move above the \citet{stern2012} demarcation at low redshifts.} 
	\epsscale{1.}
         \end{figure*}

	The sample described in this paper consists of 40 galaxies observed in three queued campaigns between November of 2012 and March of 2014 (SALT Proposals 2012-2-DC-002, 2013-1-DC-003, 2013-2-DC-003; P.I.: R. Hickox). For the purposes of exploring WISE and SDSS obscured quasar selection criteria, we chose objects in two different ways. In order to ensure maximum visibility with SALT, we restricted our possible targets to $-2^{\circ} < $ DEC $ < 2^{\circ}$ for the first campaign, and $-2^{\circ} < $ DEC $ < 0^{\circ}$ for the second campaign. Importantly, in both campaigns, all targets have $g < 22$\footnote{The SDSS magnitudes used for selection are the DR9 ``model'' AB magnitudes.} in order to increase the likelihood of obtaining a spectrum with high signal-to-noise with SALT. We further limit the sample to $g > 20$ to probe higher redshifts and avoid bright objects with existing SDSS spectroscopy. The ramifications of this optical selection criteria are described in Section \ref{sec:opticalspectra}. While the majority of the objects in our sample were chosen for optical followup due to the absence of an existing SDSS spectrum, both J0325-0032\footnote{We refer to the objects in our sample using shortened names. The full names are given in Table \ref{tab:samplespec}.} and J0924+0027 were selected so that the SALT spectrum could be compared to the SDSS spectrum, as discussed in Section \ref{sec:opticalspectra}.

\begin{deluxetable*}{lcccccccccccc}
\tabletypesize{\scriptsize}
\tablecaption{Sample Photometric Properties \label{tab:samplephot}}
\tablewidth{0pt}
\tablehead{
\colhead{SDSS Name}  & \colhead{Group}  &  \multicolumn{5}{c}{SDSS Photometry$^{a}$} & \multicolumn{4}{c}{WISE Photometry$^{b}$} & & \\ 
 &  & \colhead{$u$} & \colhead{$g$} & \colhead{$r$} & \colhead{$i$} & \colhead{$z$}
& \colhead{W1} & \colhead{W2} & \colhead{W3} & \colhead{W4} & \colhead{W1$-$W2} & \colhead{$r-\mathrm{W2}_{\mathrm{AB}}$}
}
\startdata
J032533.31--003216.4 & 1 & 21.92 & 20.79 & 19.64 & 19.56 & 18.98 & 14.59 & 12.98 & 9.56 & 6.66 & 1.60 & 3.32\\
J033820.71--004935.6 & 2 & 21.71 & 20.89 & 20.15 & 19.69 & 19.34 & 14.98 & 13.76 & 10.56 & 7.78 & 1.22 & 3.05\\
J035726.82--002724.9 & 2 & 21.60 & 20.84 & 20.44 & 19.96 & 19.73 & 14.94 & 13.96 & 10.87 & 8.39 & 0.98 & 3.14\\
J073745.20--005229.8 & 2 & 21.78 & 20.17 & 22.46 & 19.84 & 19.13 & 15.97 & 15.00 & 11.79 & --- & 0.97 & 4.12\\
J085259.35+013715.1 & 1 & 21.62 & 20.47 & 19.10 & 19.20 & 18.55 & 15.47 & 14.48 & 10.15 & 6.94 & 1.00 & 1.29\\
J092435.55+002716.4 & 1 & 21.91 & 21.41 & 20.32 & 19.46 & 18.87 & 13.18 & 12.06 & 9.30 & 6.81 & 1.12 & 4.92\\ 
J095718.06--012049.1 & 2 & 24.63 & 21.97 & 20.54 & 19.90 & 19.14 & 14.99 & 13.88 & 10.43 & 8.18 & 1.11 & 3.32\\
J100346.93--011015.9 & 1 & 22.95 & 21.91 & 20.90 & 20.38 & 19.41 & 14.25 & 12.99 & 9.67 & 6.95 & 1.27 & 4.57\\
J100711.52+014627.1 & 1 & 22.46 & 20.21 & 19.31 & 18.76 & 18.55 & 14.47 & 13.00 & 9.42 & 6.96 & 1.47 & 2.97\\
J110722.11+013336.8 & 1 & 23.51 & 21.60 & 19.96 & 19.30 & 18.82 & 14.69 & 13.74 & 9.93 & 6.89 & 0.95 & 2.88\\
J112931.47+010254.7 & 1 & 21.63 & 20.58 & 19.37 & 19.05 & 18.65 & 14.27 & 12.84 & 9.17 & 6.43 & 1.43 & 3.19\\
J113635.48+015252.9 & 1 & 22.25 & 20.85 & 19.62 & 18.84 & 18.43 & 13.63 & 12.58 & 9.60 & 6.93 & 1.04 & 3.70\\
J115158.63--004641.2 & 1 & 22.00 & 20.15 & 19.25 & 18.86 & 18.67 & 15.20 & 13.76 & 9.90 & 7.15 & 1.44 & 2.15\\ 
J125521.24--001018.2 & 1 & 21.65 & 20.56 & 19.33 & 18.94 & 18.44 & 15.01 & 13.97 & 10.18 & 7.00 & 1.03 & 2.02\\
J130500.31+005422.1 & 1 & 20.89 & 20.21 & 19.32 & 18.97 & 18.57 & 15.10 & 14.27 & 9.78 & 6.57 & 0.82 & 1.71\\
J130845.53+015542.0 & 1 & 22.35 & 21.35 & 20.12 & 19.88 & 19.40 & 13.43 & 11.99 & 9.09 & 6.88 & 1.45 & 4.80\\
J132031.04--010248.3 & 2 & 22.98 & 20.78 & 19.83 & 19.46 & 19.06 & 13.87 & 12.63 & 9.97 & 7.65 & 1.23 & 3.86\\
J132648.81--003757.6 & 2 & 22.23 & 21.54 & 21.32 & 21.10 & 20.53 & 15.77 & 14.82 & 11.12 & --- & 0.95 & 3.16\\
J133331.15--012653.3 & 1 & 21.53 & 20.41 & 19.29 & 18.63 & 18.38 & 14.88 & 14.11 & 10.04 & 6.66 & 0.77 & 1.84\\
J135423.71--000314.5 & 1 & 21.88 & 21.13 & 20.55 & 20.43 & 19.82 & 17.54 & 16.46 & 10.22 & 6.97 & 1.08 & 0.76\\
J135534.66--002206.1 & 2 & 22.79 & 21.02 & 19.62 & 19.09 & 18.28 & 13.15 & 12.16 & 9.72 & 7.53 & 0.99 & 4.12\\
J140618.16--004923.1 & 2 & 22.01 & 20.99 & 19.49 & 18.97 & 18.58 & 14.15 & 12.95 & 9.89 & 7.06 & 1.20 & 3.21\\
J140830.13--005001.6 & 2 & 22.83 & 21.93 & 20.65 & 19.73 & 19.24 & 13.97 & 12.79 & 10.04 & 8.08 & 1.18 & 4.52\\
J141724.04+010843.1 & 1 & 21.08 & 20.40 & 19.73 & 19.17 & 18.75 & 14.87 & 14.21 & 9.80 & 6.58 & 0.66 & 2.18\\
J143459.27--014432.8 & 1 & 21.95 & 20.40 & 19.40 & 18.82 & 18.48 & 15.20 & 14.02 & 9.71 & 6.62 & 1.18 & 2.04\\
J144006.46--011624.7 & 1 & 21.36 & 20.33 & 19.24 & 18.84 & 18.43 & 13.69 & 12.40 & 9.29 & 6.78 & 1.28 & 3.50\\
J144625.94--015721.9 & 2 & 22.18 & 21.50 & 20.46 & 19.94 & 19.45 & 14.11 & 13.30 & 10.21 & 8.24 & 0.81 & 3.82\\
J150539.97+013433.9 & 1 & 21.69 & 20.44 & 19.28 & 18.88 & 18.54 & 15.02 & 13.89 & 9.62 & 6.93 & 1.13 & 2.05\\
J152736.35--001007.7 & 2 & 21.02 & 20.89 & 20.12 & 19.85 & 18.90 & 14.44 & 13.45 & 11.65 & --- & 0.99 & 3.33\\
J153846.30+012951.9 & 1 & 23.55 & 21.25 & 20.25 & 19.39 & 19.35 & 15.13 & 13.57 & 9.55 & 6.89 & 1.56 & 3.34\\
J154503.94--010010.4 & 1 & 21.67 & 20.86 & 19.38 & 18.70 & 18.43 & 13.51 & 12.10 & 9.13 & 6.80 & 1.41 & 3.94\\
J155421.94+001115.0 & 1 & 22.72 & 21.09 & 19.92 & 18.85 & 18.54 & 14.49 & 13.67 & 9.95 & 6.69 & 0.82 & 2.91\\
J160903.75--000426.2 & 2 & 22.65 & 20.99 & 20.12 & 19.63 & 19.18 & 14.88 & 13.73 & 10.22 & --- & 1.15 & 3.05\\
J160928.56--013344.1 & 2 & 25.86 & 21.62 & 20.09 & 19.29 & 18.73 & 14.56 & 13.57 & 10.16 & 7.70 & 0.99 & 3.18\\
J163445.76--010808.6 & 2 & 21.01 & 20.46 & 20.15 & 19.76 & 19.52 & 14.41 & 13.39 & 10.90 & --- & 1.02 & 3.42\\
J172446.04--003833.4 & 2 & 21.74 & 20.62 & 19.48 & 19.00 & 18.26 & 13.45 & 12.39 & 9.54 & 7.16 & 1.07 & 3.75\\
J173114.67--003708.1 & 2 & 22.09 & 21.25 & 24.81 & 24.37 & 20.05 & 16.12 & 14.90 & 12.14 & --- & 1.22 & 6.57\\
J180853.93--001650.6 & 2 & 21.53 & 20.65 & 19.88 & 19.52 & 19.26 & 14.14 & 13.04 & 10.13 & 7.61 & 1.11 & 3.50\\
J194745.03--003603.8 & 2 & 22.76 & 21.73 & 21.20 & 20.98 & 20.33 & 15.42 & 14.42 & 10.93 & 8.31 & 1.00 & 3.44\\
J202952.16--010805.3 & 2 & 21.89 & 21.02 & 20.08 & 19.51 & 19.26 & 14.67 & 13.31 & 10.48 & --- & 1.36 & 3.44
\enddata
\tablenotetext{a}{DR9 ``model'' AB magnitudes.}
\tablenotetext{a}{Where the WISE ext\_flg value indicates that the objects are unresolved, we use the ``profile-fitting'' magnitudes (\textit{w*mpro}). For those where the object is resolved, we use the ``standard'' aperture magnitudes (\textit{w*mag}). In both cases, the photometry is given in Vega magnitudes.}
\end{deluxetable*}

	Our first selection group consists of all objects with SDSS and WISE coverage with WISE color $\mathrm{W1}- \mathrm{W2} > 0.7$\footnote{Using the WISE ext\_flg value, 39 objects are unresolved in the WISE photometry, and for these we use the ``profile-fitting'' photometric magnitudes taken from the WISE All-Sky Source Catalog. For  J1609-0004, we use the ``standard'' aperture magnitudes. For any other samples discussed in this paper, we use the corresponding magnitudes based on the ext\_flg values. All WISE photometry is given in Vega magnitudes, unless specified.}, a slightly more relaxed criteria for selecting quasars from what is presented in \citet{stern2012}. In practice, while only one object, J1417+0108, has $\mathrm{W1}- \mathrm{W2} = 0.73$, the rest have $\mathrm{W1}- \mathrm{W2} > 0.8$. In order to select objects with the brightest mid-IR emission, we further restricted the possible candidates to only those objects with $7 \ge \mathrm{W4} \ge 6.5$.  These relatively rare IR-bright objects are comparable to the \emph{Spitzer} 24 $\mu$m-bright objects explored in \citet{lacy2013}. We acquired SALT spectroscopy for 21 objects selected under these criteria, which we refer to as ``Group 1.''
	
	For the second campaign, we started with only those objects with SDSS and WISE coverage with WISE color $\mathrm{W1}- \mathrm{W2} > 0.8$, and $\mathrm{W4} \ge 7.0$. We further selected objects with $r_{\mathrm{AB}}-\mathrm{W2}_{\mathrm{AB}} > 3.1$, the criteria introduced in \citet{hickox2007} that targets obscured quasars. We acquired SALT spectroscopy for 19 objects selected under these criteria, which we refer to as ``Group 2.'' If we only explore the overlapping range in DEC used in selecting our targets for both Group 1 and Group 2 ($-2^{\circ} < $ DEC $ < 0^{\circ}$) across all RAs, we find 144 objects qualify as Group 1 objects, while 1270 objects qualify as Group 2 objects, due largely to the relaxation of the W4 photometry criteria for Group 2. Additionally, in this same DEC range, there are 77 objects that would be selected as Group 1 objects with $g > 22$, while there would be 8166 Group 2 objects with $g > 22$, which represents a significant number of candidate obscured quasars that are missed due to our optical photometric cuts. For the full DEC range used to find Group 1 objects ($-2^{\circ} < $ DEC $ < 2^{\circ}$), there are 277 objects that would be selected under the Group 1 criteria, while an additional 145 objects would be selected under these criteria but with $g > 22$.
	
	We plot the WISE W1$-$W2 and W2$-$W3 colors for both objects in both Group 1 and 2 in Figure \ref{fig:wisecolorcolor} along with AGN selection criteria taken from the literature. We also plot the SDSS-selected Type II quasars (with WISE W1, W2, and W3 detections with SNR $> 3.0$) from \citet{reyes2008} for comparison. The objects in our sample cover a wide range in colors. To indicate the WISE colors of star-forming galaxies as a function of redshift from $z = 0-2$, we also plot tracks showing the \citet{assef2010} star-forming (``Sbc'') and irregular (``Im'') templates, as well as SEDs for the star-forming galaxies M82 and Arp220 from \citet{polletta2007}. We further discuss the implications of our selection criteria in Section \ref{sec:agnselection}. The objects in our sample do not have Chandra coverage with any instrument, and while there is XMM-Newton coverage for fields containing the objects J1505+0134 and J1129+0102\footnote{The XMM-Newton observations are Obs-ID:0021540101 (30 ks), 0021540501 (20ks), 0723800101 (82ks), and 0723800201 (89ks) for the field containing J1505+0134, and Obs-ID: 0305750601 (6 ks) for the field containing J1129+0102.}, the objects were not detected. We also note that 8 of our 45 objects have detections within $2''$ in the VLA Faint Images of the Radio Sky at Twenty-Centimeters (FIRST) survey \citep{becker1995}: J1007+0146 (1.27 mJy integrated 21cm flux, 1.24 mJy/beam peak flux), J0852+0137 (3.96 mJy, 3.59 mJy/bm), J0957-0120 (2.78 mJy, 2.20 mJy/bm), J1151-0046 (2.85 mJy, 3.16 mJy/bm), J1333-0126 (1.25 mJy, 1.51 mJy/bm), J1406-0049 (3.20 mJy, 3.16 mJy/bm), J1417+0108 (1.38 mJy, 1.64 mJy/bm), and J1554+0011 (6.15 mJy, 5.78 mJy/bm). These detections could be caused by both star formation as well as AGN activity \citep{ivezic2002}.
		
	The objects were selected using photometry from SDSS Data Release 9 \citep{ahn2012} as well as the WISE All-Sky Source Catalogue. Recently, the WISE team released the AllWISE Source Catalog \citep[][made public November 2013]{cutri2013}, which combined three WISE surveys: the 4-band Cryogenic Survey (the primary WISE mission that covered the full sky 1.2 times between January 2010 and August 2010), the 3-Band Cryogenic Survey (a survey using only the first three WISE wavebands, which covered 30\% of the sky between January 2010 and August 2010), and the NEO-WISE post-cryogenic survey (a survey using the first two WISE wavebands, which covered 70\% of the sky between September 2010 and February 2011). Most importantly for the purposes of the current study, the AllWISE Data includes updated photometry more sensitive in the W1 and W2 bands, with updated 5$\sigma$ point-source sensitivities at 0.054 and 0.071 mJy, respectively \citep{cutri2013}. For the remainder of this paper, we will report the updated, more accurate AllWISE photometry for the objects in our sample. The average fluxes in the W1 and W2 band for the full selection of objects are 600 and 900 microJanskys, respectively. For the majority of the objects, the updated AllWISE photometry does not affect whether or not these objects would be selected, although there are a few objects which would move into or out of our selection boxes. The Group 1 object J1417+0108 has updated photometry with WISE color W1$-$W2 $< 0.7$, which does not satisfy our initial color cut. Similarly, the Group 1 objects J1151-0046 and J1129+0102 have W4-band photometry that excludes them from selection under Group 1 criteria. Finally, J0338-0049 and J1609-0004, two Group 2 objects, have updated color $r_{\mathrm{AB}}-\mathrm{W2}_{\mathrm{AB}} < 3.1$. We further discuss these targets in Section \ref{sec:individualobjects}. For the W1, W2, and W3 photometry used in our selection, all but one of our objects has AllWISE signal-to-noise ratios (SNR) on the photometry above 3.0 (J1609-0004 has SNR$_{\mathrm{W1}} = 2.9$). We only use the observed photometry in the W4 band if the object was observed with SNR $> 3.0$. We report the object identifiers, group number, SDSS and WISE AllWISE Source Catalogue photometry, and the IR and optical-to-IR colors used to select the objects in Table \ref{tab:samplephot}.

	In order to better explore how the change in the WISE photometry affected the selection of targets, we explored all of the objects that would be selected as Group 1 and Group 2 objects in the range $40^{\circ} < $ RA $ < 185^{\circ}$, $-2^{\circ} < $ DEC $ < 0^{\circ}$ with both WISE All-Sky and AllWISE photometry. For Group 1, 49 total objects would be selected using both the All-Sky and AllWISE photometry, and an additional 5 objects moved into the selection criteria and 14 moved out of the selection criteria based on the updated AllWISE photometry, primarily because of updated AllWISE photometry with $\mathrm{W4} > 7$ or $ \mathrm{W4} < 6.5$ for individual objects. For Group 2, 323 total objects would be selected using both the WISE All-Sky and AllWISE photometry, and an additional 56 objects moved into the selection criteria and 28 moved out of the selection criteria based on the updated AllWISE photometry, due to the larger number of criteria used for selecting an object in Group 2. These results demonstrate the increased sensitivity of the updated AllWISE data, and we reiterate that except where noted above, the objects in our sample would be selected to be a part of Group 1 or Group 2 using either the AllWISE or WISE All-Sky photometry. 

	The full sample of objects were observed using the Robert Stobie Spectrograph \citep[RSS,][]{kobulnicky2003,smith2006} on SALT in long-slit mode, with a 1\secpoint5 slit and the PG0900 grating, which provides a spectral resolution of 5 \AA\, at a central wavelength of 5000 \AA\, (in this observational set-up, the spectral range was $3600 - 6000$ \AA). The seeing was measured using stars observed in the acquisition images taken just before the observations, and was typically $\sim 2''$ for the runs. The slit positions were chosen to coincide with a nearby bright source for ease of object acquisition. While the majority of the objects comprising the sample were observed for 2250 seconds, for a small number of objects with brighter or dimmer optical photometry, the exposure time was changed accordingly, leading to a full range of observing times between 1950 and 3000 seconds. Data reduction was performed using standard IRAF scripts\footnote{IRAF is distributed by the National Optical Astronomy Observatory, which is operated by the Association of Universities for Research in Astronomy (AURA), Inc., under cooperative agreement with the National Science Foundation.}. The data were gain-corrected, bias-corrected, and mosaicked with the SALT reduction pipeline. We subsequently flat-fielded the data, applied a wavelength solution using arc lamp spectra, and then used the three observations made for each target for cosmic ray removal. In our final step, the two-dimensional spectra were background subtracted and combined using a median combine. The spectra were extracted using a $2.7''$ aperture centered on the target. 
	
	We also observed standard stars in order to perform relative flux calibration on the data and correct for the sensitivity of the SALT CCD. Unfortunately, due to the fixed nature of the primary mirror at the SALT telescope and the strong variation in effective aperture with time and source position, it is not possible to perform absolute flux calibration using the SALT data alone. We will report emission line flux ratios for a subsample of the objects in Section \ref{sec:linediagnostics}. We report the details of the spectroscopic observations, including observation date and total exposure time, in Table \ref{tab:samplespec}.

\section{Optical Spectra and Redshift Distribution}
\label{sec:opticalspectra}

	\begin{figure*}[htbp]
	\epsscale{1.05} 
	\plotone{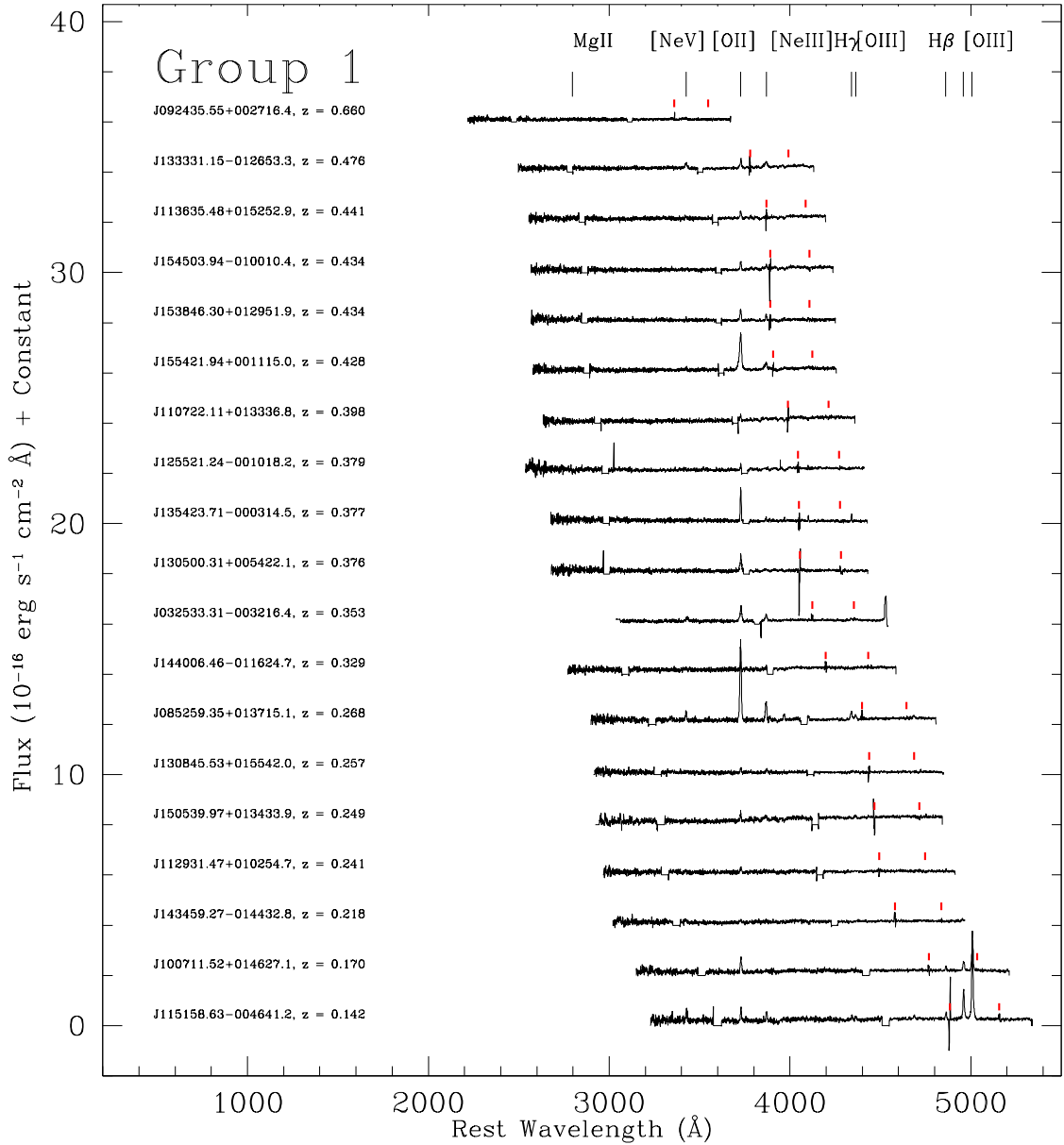}
	\caption{
	\label{fig:group1redshifts} Rest-frame spectra for objects in Group 1. We plot only those objects with redshift determinations that could be made from identifying one or more emission lines in the observed frame spectrum. We also plot the positions of the strongest optical lines used for determining redshift above the spectra, and indicate the two strongest sky features in each spectrum with red vertical lines.} 
	\epsscale{1.}
         \end{figure*}

	\begin{figure*}[htbp]
	\epsscale{1.05} 
	\plotone{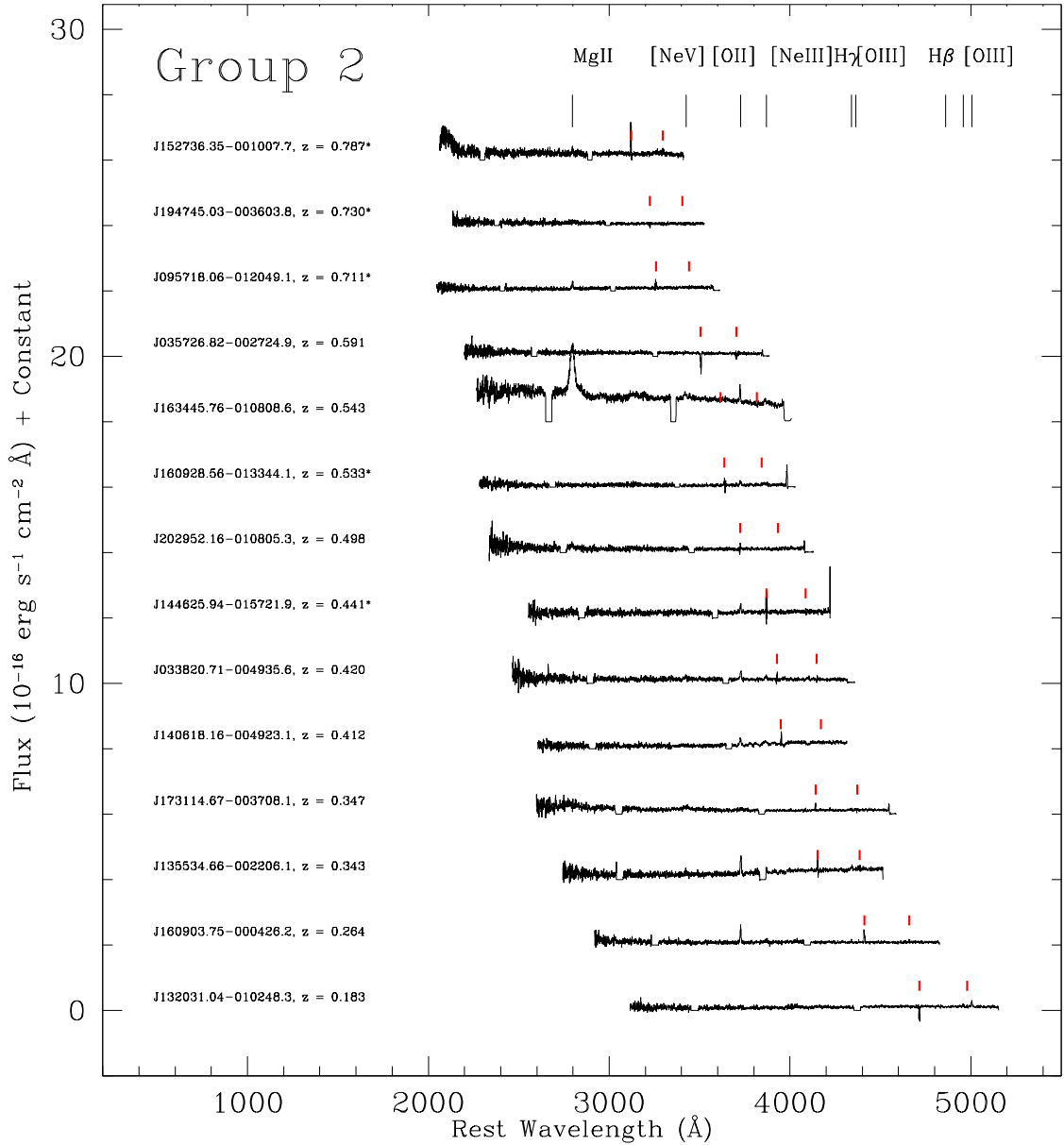}
	\caption{
	\label{fig:group2redshifts} Same as Figure \ref{fig:group1redshifts}, but for those objects from Group 2. We also indicate with an asterisk which objects have redshift determinations from a single emission line.} 
	\epsscale{1.}
         \end{figure*}

	\begin{figure*}[htbp]
	\epsscale{1.05} 
	\plotone{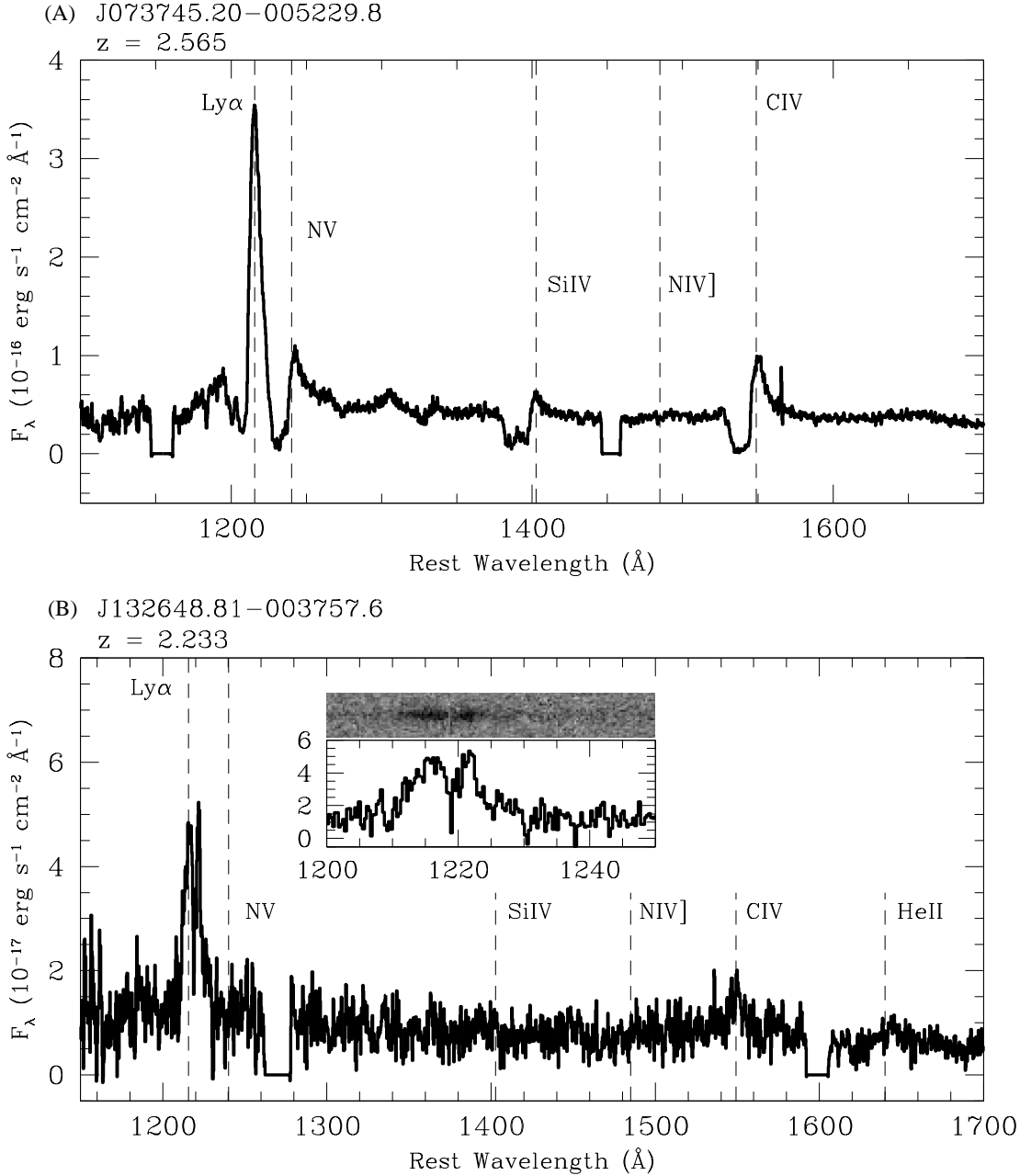}
	\caption{
	\label{fig:hi_z_figures} Rest-frame spectra for the two high-redshift objects J0737-0052 (A, $z = 2.565$) and J1326-0037 (B, $z = 2.233$), both from Group 2. For J0737-0052, the presence of Ly$\alpha$, NV$\lambda$1240, and CIV$\lambda$1548 in emission is indicative of the presence of an AGN in this object. The broad absorption features seen to the blue of each strong emission line indicates that this object is most likely a broad absorption line quasar. The spectrum for J1326-0037 shows Ly$\alpha$ and CIV$\lambda$1548 in emission, and in the inset, we plot a zoom on the Ly$\alpha$ line as well as the 2D spectrum.} 
	\epsscale{1.}
         \end{figure*}

	\begin{figure*}[htbp]
	\epsscale{1.05} 
	\plotone{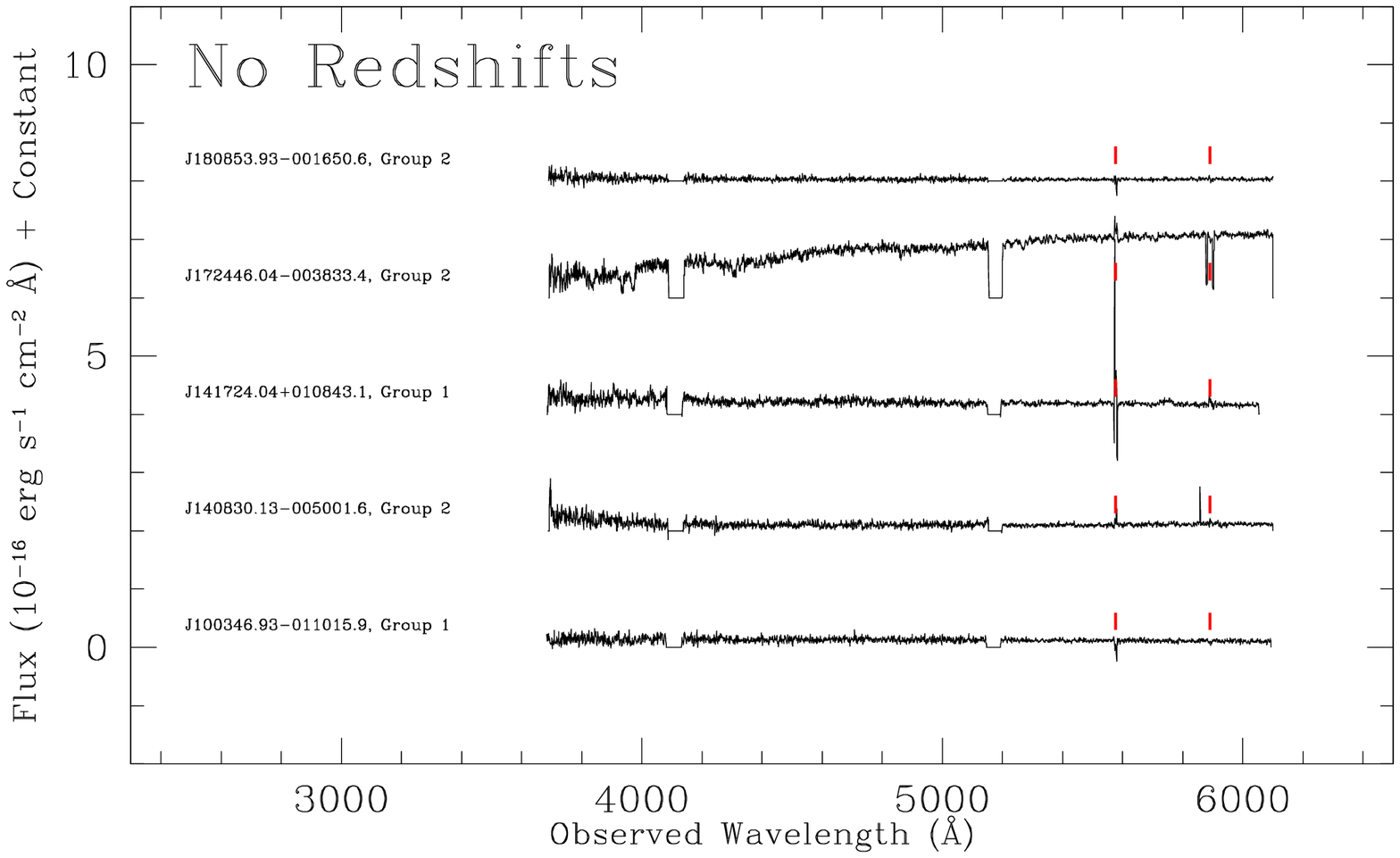}
	\caption{
	\label{fig:group3redshifts} Observed-frame spectra for the objects in our sample without reliable redshift estimates. We indicate the group that each object belongs to on the figure.} 
	\epsscale{1.}
         \end{figure*} 
                 
	With the assembled, final, reduced spectra, our first task was estimating the spectroscopic redshift of each object, and initial identification of emission lines was done by eye for each object. The strongest optical lines we used to constrain the redshift were MgII$\lambda$2798, [NeV]$\lambda$3427, [OII]$\lambda\lambda$3726,3729, [NeIII]$\lambda$3869, H$\beta$, and [OIII]$\lambda\lambda$4959,5007. Additionally, the two highest-redshift objects in our sample were identified using strong UV emission lines. For the majority of the objects (29/40), we were able to identify two or more emission lines in the spectrum which we used to constrain the redshift, while for J1255-0010 there was only one strong emission line which we identified as [OII]$\lambda\lambda$3726,3729. Due to a strong observed 4000 \AA\, break to the red of the emission line in the spectrum for this object, we consider the redshift estimate for J1255-0010 to be confident. For 5 other objects that have a single emission line, we used the fact that there were no other strong emission lines in the spectrum to infer that the single line was most likely MgII$\lambda$2798, which was also observed to be broad in one of the objects (see Section \ref{sec:unobscured} for a discussion of this object). For the remaining objects (5/40), there were no features in the spectrum allowing for accurate redshift identification. One of these objects, J1417+0108, does have a FIRST radio detection, however. In light of this, while we calculated the statistical properties of the sample using all of the estimated redshifts, we will also present statistics calculated without the 5 objects with a single emission feature. 

	Redshifts were estimated from the centroids measured using the IRAF task \emph{splot} to fit Gaussian curves to the emission lines in each spectrum. In order to characterize the systematic error on our redshift estimates introduced due to wavelength calibration we compared SALT RSS spectra observed in a different campaign \citep[as described in][]{hainline2013} to the measured SDSS spectra for these objects, and found that the average difference was on the order of $\sim0.3$\AA. We can also compare the emission line measurements for the two objects in our sample for which SDSS DR9 spectra were taken: J0325-0032 ($z = 0.353$) and J0924+0027 ($z = 0.660$). For J0325-0032, the best-fit centroids for the strongest lines [OII]$\lambda\lambda$3726,3729 and [NeIII]$\lambda$3869 agree to within $\sim 3$ \AA, while for J0924+0027 the best-fit centroids for the more broad, low S/N emission line [NeV]$\lambda$3427 agrees to within $\sim 8$ \AA. 

\begin{deluxetable*}{lcccccccc}
\tabletypesize{\scriptsize}
\tablecaption{Sample Spectroscopic Properties \label{tab:samplespec}}
\tablewidth{0pt}
\tablehead{
\colhead{SDSS Name} & \colhead{Group} & \colhead{Obs. Date} & \colhead{Exp. Time (s)} & \colhead{$z_{\mathrm{spec}}$} & \colhead{$z_{\mathrm{phot}}$\tablenotemark{a}} & \colhead{[NeIII] / [OII]\tablenotemark{b}} & \colhead{$^{0.0}(g-z)$\tablenotemark{c}} & \colhead{log($L_{8\mu\mathrm{m}}$ / erg s$^{-1}$)\tablenotemark{d}}
}
\startdata

J032533.31--003216.4 & 1 & Nov 2012 & 2610.6 & 0.353 & 0.140 & $0.44 \pm 0.02$ & $0.77 \pm 0.07$ & 44.79\\
J033820.71--004935.6 & 2 & Oct 2013 & 2250.6 & 0.420 & 0.280 & $0.42 \pm 0.07$ & $0.95 \pm 0.16$ & 44.61\\
J035726.82--002724.9 & 2 & Oct 2013 & 2250.6 & 0.591 & 0.580 & - & - & 44.90\\ 
J073745.20--005229.8\tablenotemark{e} & 2 & Nov 2013 & 1950.6 & 2.565 & 2.420 & - & - & 46.58\\ 
J085259.35+013715.1 & 1 & Mar 2013  & 2250.6 & 0.268 & 0.160 & $0.23 \pm 0.00$ & $1.13 \pm 0.04$ & 44.17\\
J092435.55+002716.4 & 1 & Feb 2013  & 2820.6 & 0.660 & 0.700 & - & - & 45.66\\
J095718.06--012049.1 & 2 & Mar 2014 & 2620.6 & 0.711 & 0.660 & - & - & 45.33\\
J100346.93--011015.9 & 1 & Mar 2013 & 2250.6 & - & 0.340 & - & - & 44.71\tablenotemark{f}\\
J100711.52+014627.1 & 1 & Mar 2013  & 2250.6 & 0.170 & 0.220 & $0.11 \pm 0.02$ & $1.36 \pm 0.07$ & 44.01\\
J110722.11+013336.8 & 1 & Apr 2013   & 2250.6 & 0.398 & 0.390 & $0.31 \pm 0.03$ & $1.60 \pm 0.16$ & 44.77\\
J112931.47+010254.7 & 1 & Mar 2013  & 2250.6 & 0.241 & 0.310 & $0.45 \pm 0.09$ & $1.31 \pm 0.10$ & 44.49\\
J113635.48+015252.9 & 1 & Mar 2013  & 2250.6 & 0.441 & 0.470 & $< 0.89$ & $2.17 \pm 0.17$ & 45.05\\
J115158.63--004641.2 & 1 & Mar 2013 & 2250.6 & 0.142 & 0.050 & $0.83 \pm 0.05$ & $1.14 \pm 0.02$ & 43.60\\
J125521.24--001018.2 & 1 & Mar 2013 & 2550.6 & 0.379 & 0.270 & $< 5.26$ & $1.31 \pm 0.12$ & 44.61\\
J130500.31+005422.1 & 1 & Mar 2013 & 2250.6 & 0.376 & 0.310 & $0.08 \pm 0.01$ & $1.28 \pm 0.10$ & 44.74\\
J130845.53+015542.0 & 1 & Mar 2013 & 2250.6 & 0.257 & 0.020 & $0.88 \pm 0.10$ & $1.23 \pm 0.10$ & 44.65\\ 
J132031.04--010248.3 & 2 & May 2013 & 2250.6 & 0.183 & 0.380 & $< 1.02$ & $1.28 \pm 0.07$ & 43.96\\
J132648.81--003757.6 & 2 & May 2013 & 2250.6 & 2.233 & 2.550 & - & - & 46.80\\
J133331.15--012653.3 & 1 & Feb 2013 & 3000.6 & 0.476 & 0.220 & $0.73 \pm 0.03$ & $0.75 \pm 0.15$ & 44.96\\
J135423.71--000314.5 & 1 &  Mar 2013 & 2250.6 & 0.377 & 0.340 & $0.10 \pm 0.01$ & $0.96 \pm 0.22$ & 44.51\\
J135534.66--002206.1 & 2 & May 2013 & 2250.6 & 0.343 & 0.260 & - & - & 44.74\\
J140618.16--004923.1 & 2 & May 2013 & 2250.6 & 0.412 & 0.340 & $0.44 \pm 0.06$ & $1.13 \pm 0.09$ & 44.85\\
J140830.13--005001.6 & 2 & Jun 2013  & 2250.6 & - & 0.640 & - & - & 45.32\tablenotemark{f}\\
J141724.04+010843.1 & 1 & Mar 2013  & 2250.6 & - & 0.060 & - & - & 42.70\tablenotemark{f}\\
J143459.27--014432.8 & 1 &  Mar 2013 & 2250.6 & 0.218 & 0.220 & $0.64 \pm 0.18$ & $1.04 \pm 0.10$ & 44.10\\
J144006.46--011624.7 & 1 & Mar 2013  & 2250.6 & 0.329 & 0.170 & - & - & 44.83\\
J144625.94--015721.9 & 2 & May 2013 & 2430.6 & 0.441 & 0.700 & - & - & 44.81\\
J150539.97+013433.9 & 1 & Mar 2013  & 2250.6 & 0.249 & 0.240 & $0.48 \pm 0.07$ & $1.25 \pm 0.05$ & 44.30\\
J152736.35--001007.7 & 2 & May 2013 & 2250.6 & 0.787 & 0.970 & - & - & 44.87\\
J153846.30+012951.9 & 1 & Apr 2013  & 2250.6 & 0.434 & 0.140 & $0.47 \pm 0.02$ & $0.98 \pm 0.19$ & 45.04\\
J154503.94--010010.4 & 1 & Mar 2013  & 2250.6 & 0.434 & 0.440 & $0.41 \pm 0.03$ & $1.28 \pm 0.13$ & 45.22\\
J155421.94+001115.0 & 1 & Apr 2013  & 2250.6 & 0.428 & 0.480 & $0.40 \pm 0.07$ & $1.67 \pm 0.17$ & 44.86\\
J160903.75--000426.2 & 2 &  Aug 2013 & 1950.6 & 0.264 & 0.210 & $< 0.39$ & $1.41 \pm 0.27$ & 44.17\\ 
J160928.56--013344.1 & 2 & Mar 2014 & 2100.6 & 0.533 & 0.460 & - & - & 45.06\\
J163445.76--010808.6\tablenotemark{e} & 2 & Aug 2013 & 2250.6 & 0.543 & 0.120 & - & - & 44.78\\
J172446.04--003833.4 & 2 & Aug 2013 & 2250.6 & - & 0.900 & - & - & 45.96\tablenotemark{f}\\
J173114.67--003708.1 & 2 & Aug 2013 & 2250.6 & 0.347 & 0.440 & - & - & 43.77\\
J180853.93--001650.6 & 2 & Jul 2013  & 2250.6 & - & 0.410 & - & - & 44.76\tablenotemark{f}\\
J194745.03--003603.8 & 2 & Aug 2013 & 1950.6 & 0.730 & 0.360 & - & - & 45.17\\
J202952.16--010805.3 & 2 & Sep 2013 & 2250.6 & 0.498 & 0.440 & - & - & 44.83
\enddata
\tablenotetext{a}{These photometric redshift values were estimated using SED modeling as described in Section \ref{sec:sedmodeling}.}
\tablenotetext{b}{Flux ratio between [NeIII]$\lambda$3869 and [OII]$\lambda\lambda$3726+3729 emission features.}
\tablenotetext{c}{Rest-frame $g-z$ color.}
\tablenotetext{d}{Luminosity at 8$\mu$m estimated from the WISE photometry.}
\tablenotetext{e}{These objects have broad emission lines in their observed spectra, and we classify them as Type I quasars.}
\tablenotetext{f}{These values for $L_{8\mu\mathrm{m}}$ were derived using $z_{\mathrm{phot}}$.}
\end{deluxetable*}        
	
	In Group 1, we identified the redshift of 19/21 (90\%) of the objects, and in Group 2, we identified the redshift of 16/19 (84\%) of the objects (12/19, 63\%, were identified with more than one emission line). For those spectra where we were able to estimate the redshift we present the rest-frame spectra in Figures \ref{fig:group1redshifts} (Group 1) and \ref{fig:group2redshifts} (Group 2). In these figures, we note those spectra with only one identified emission line with an asterisk. In Figure \ref{fig:hi_z_figures} we highlight our two highest redshift sources, J0737-0052 ($z = 2.565$) and J1326-0037 ($z = 2.233$), both of which are Group 2 objects. Finally, we plot the 5 spectra in the observed frame without reliable redshift information in Figure \ref{fig:group3redshifts}. We list the spectroscopic redshifts for the sample in Table \ref{tab:samplespec}. 

\begin{figure}[htbp]
	\epsscale{1.2}          
	\plotone{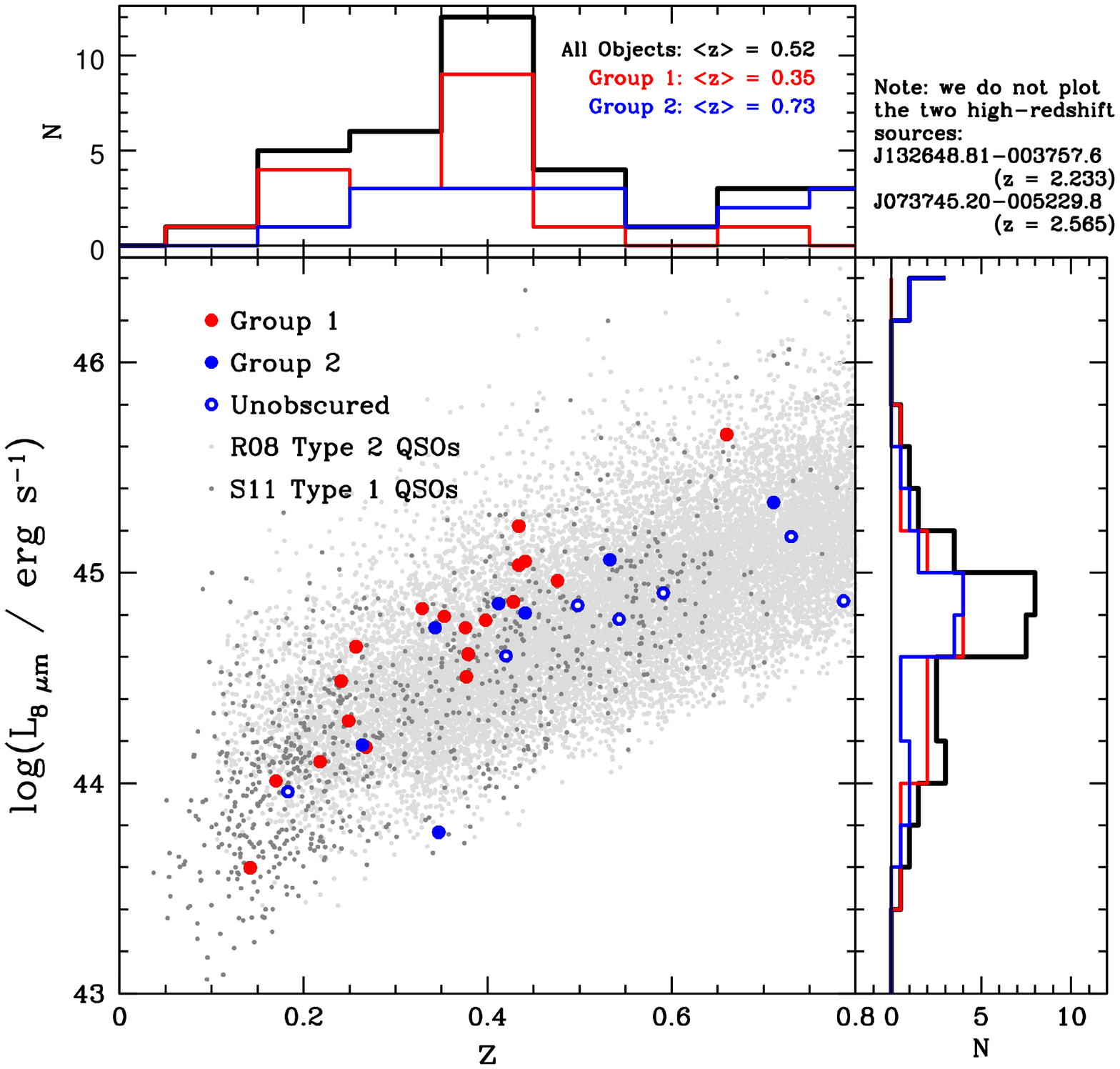}
 	\caption{
	\label{fig:lum_vs_z} WISE 8$\mu$m Luminosity plotted against spectroscopic redshift for the objects in our sample. We also show the IR luminosities and redshifts for the SDSS Type I quasar sample from \citet{shen2011} in light grey, and the SDSS Type II quasar sample from \citet{reyes2008} in dark grey. For both of these comparison samples, we only plot 8-micron luminosities for objects with WISE W1, W2, and W3 SNR $> 3.0$. Our SALT sample is plotted with red (Group 1) and blue (Group 2) points (with unobscured objects from the SED fitting shown with open circles), and we also show histograms of the distributions on the top and right sides of the figure. We truncate the plot at $z\sim0.8$ and do not show the two highest redshift objects in our sample. The objects in our SALT sample have IR luminosities that span the full range observed for both Type I and II quasars, although, at a given redshift, our selection criteria does preferentially choose the most IR-luminous objects as compared to those selected optically.} 
	\epsscale{1.}
         \end{figure}

	One of the fundamental properties of our WISE-selected sources is their redshift distribution, which we plot for our sample in the top panel of Figure \ref{fig:lum_vs_z}. We plot the redshift distribution for all of the objects (excluding the two highest-redshift objects) in black, the AGNs from Group 1 in red, and the AGNs from Group 2 in blue. The full redshift distribution has an average of $\langle z_{all} \rangle = 0.52$, Group 1 has an average of $\langle z_{1} \rangle = 0.35$, and Group 2 has an average of $\langle z_{2} \rangle = 0.73$ (the average redshift becomes $\langle z_{2} \rangle = 0.49$ without including the two $z > 2$ objects). We performed a Kolmogorov-Smirnov two-sample test on the redshift distributions from the two groups and calculated the K-S statistic $P = 0.01$, indicating that these two samples were drawn from different distributions. We also performed the test without the 5 objects from Group 2 with only one emission feature, and calculated $P = 0.16$, however. While this is highly dependent on the objects in our sample where we only identify one emission feature, our results indicate that by targeting objects with fainter W4 magnitudes, the distribution shifts to higher redshifts. This result is not surprising given the significantly shallower depth of WISE observations in the W4 band as compared to the other three photometric bands. By choosing only those objects that are brightest in W4, we preferentially select objects at lower redshift.

	We can interpret our redshift distribution in light of the results on WISE AGN selection described in the literature. \citet{assef2013} describe the spectroscopic redshift distribution for a sample of WISE-selected AGNs ($\mathrm{W1}- \mathrm{W2} \ge 0.8$) in the Bo\"{o}tes field, and demonstrate that the distribution is double-peaked for those objects in their sample with $\mathrm{W2} < 15.73$. The main peak is between $1 < z < 2$, while there is a smaller peak at $z \sim 0.25 - 0.5$. The authors propose that the WISE sensitivity results in recovering obscured AGNs at lower redshifts, where emission from these objects dominate the W1 and W2 photometric bands. As the redshift increases, the W1$-$W2 color criteria is less efficient at selecting obscured quasars since the portion of the SED probed by the W1 and W2 photometric bands moves to shorter wavelengths where the obscured AGN emission is not as dominant compared to host galaxy emission. This effect leads to a minimum in the distribution at $z \sim 0.75$, although there is an increase beyond that due to a larger comoving volume being observed. The distribution of redshifts for our SALT targets is very similar to the peak at $z \sim 0.25 - 0.5$ from the \citet{assef2013} distribution. The vast majority of our objects have $z < 0.75$, and the two bright $z \sim 2$ AGNs also agree with the higher-$z$ tail of the \citeauthor{assef2013} sample. 
	
	As we probed objects with fainter W4 magnitudes, between Groups 1 and 2, we found that our distribution moved to higher redshifts, but the lack of objects at $1 < z < 2$ in our sample is most likely a result of our initial optical selection criteria of $20 < g < 22$. For obscured quasars, the $g$-band probes host galaxy emission, and so this cut limits our galaxy selection to lower redshifts. To test this assumption, we explored a large catalogue of $\sim 4000$ galaxies with photometric redshifts and SDSS identifications from the Bo\"{o}tes field. For all objects with $20 < g < 22$, the average (median) photometric redshift is only $0.43 (0.37) \pm 0.24$, demonstrating the impact a SDSS $g$ band magnitude cut can have at selecting high-redshift galaxies. 
	
	These redshifts, along with the WISE photometry in the W2, W3, and W4 bands, can be used to estimate an AGN IR luminosity. Following the analysis presented in \citet{hainline2013}, we used the WISE photometry to estimate a luminosity at rest-frame 8$\mu$m ($L_{8 \mu \mathrm{m}}$). For this estimate, we modeled the AGN mid-IR emission with a power law and do not account for the individual filter response functions, although, based on the WISE colors for these objects, any flux corrections would be on the order of a few percent \citep{wright2010}. We also assume that the flux is measured at the central wavelength for each filter. In Section \ref{sec:sedmodeling}, we describe the use of galaxy and AGN templates to model the broad-band SEDs for these objects, and we can also use the best-fit AGN template to estimate the 8$\mu$m AGN luminosity for each object in our sample. We find that luminosities measured from the best-fit templates agree with those estimated from the WISE photometry (and presented in Table \ref{tab:samplespec}) to within 0.2 dex. Excluding the two high-redshift sources, the objects in our sample have a range in 8$\mu$m luminosity of $\log(L_{8 \mu \mathrm{m}}) = 44.0 - 45.0$, with an average (median) of $\langle \log(L_{8 \mu \mathrm{m}}) \rangle = 44.9 (44.8)$. In order to compare the luminosities to those from quasars in the literature, we plot the 8$\mu$m luminosity against the redshift for our sample in Figure \ref{fig:lum_vs_z}. For comparison, we also plot the estimated 8$\mu$m luminosities from the SDSS Type II quasar sample (with W1, W2, W3, SNR $> 3.0$) from \citet{reyes2008} ($\langle z \rangle = 0.324$, $z_{\mathrm{median}} = 0.279$), with an average (median) of $\langle \log(L_{8 \mu \mathrm{m}}) \rangle = 44.7 (44.3)$, as well as the estimated 8$\mu$m luminosities for the SDSS Type I quasar sample (with W1, W2, W3, SNR $> 3.0$) from \citet{shen2011} (for the quasars at $z < 0.84$, $\langle z \rangle = 0.519$, $z_{\mathrm{median}} = 0.529$), with an average (median) of $\langle \log(L_{8 \mu \mathrm{m}}) \rangle = 45.0 (44.8)$ (we restrict the objects used in this average to those with $z < 0.84$, the upper limit on the redshifts from the \citet{reyes2008} sample). Based on these results, if the mid-IR is dominated by AGN emission for the objects in our sample, these AGNs would lie in the quasar regime. Notice, however, that at each redshift, the objects that comprise our sample are at systematically higher IR luminosities compared to the optically selected sample, as would be expected from WISE selection. 
	
\section{Emission Line Diagnostics}
\label{sec:linediagnostics}

	\begin{figure}[htbp]
	\epsscale{1.2} 
	\plotone{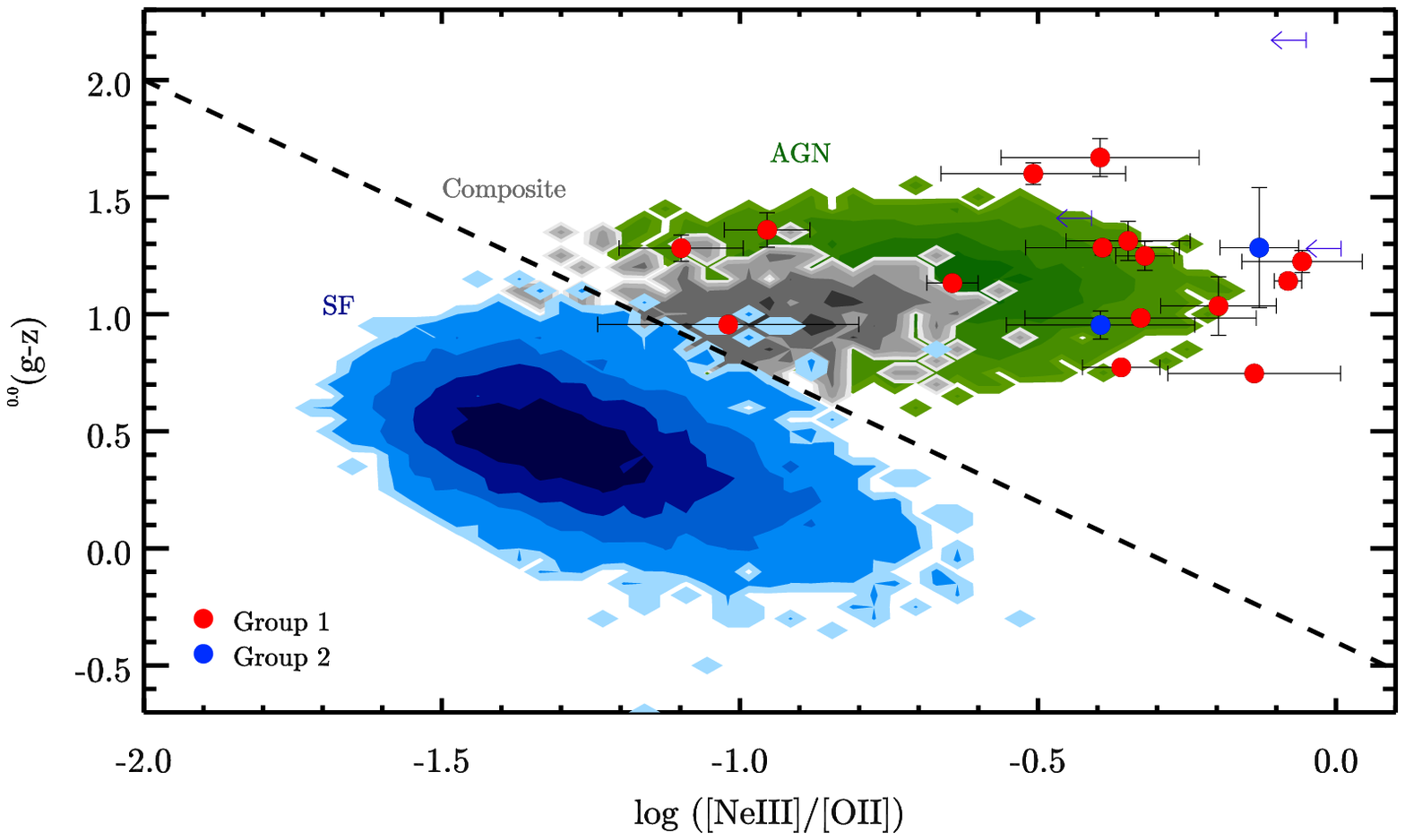}
	\caption{
	\label{fig:tbtdiagram} Rest-frame $g-z$ color plotted against log([NeIII]$\lambda$3869 / [OII]$\lambda\lambda$3726+3729) \citep[``TBT Diagram'',][]{trouille2011}. The objects in our sample with significant [NeIII]$\lambda$3869 and [OII]$\lambda\lambda$3726+3729 emission line flux measurements are plotted with red (Group 1) and blue (Group 2) circles. We also plot SDSS galaxies from \citet{trouille2011} (those with BPT and TBT emission lines detected with S/N $ > 5$) with colored contours. In green, we plot contours for the SDSS BPT AGNs, in blue we plot the SDSS BPT star-forming galaxies, and in grey we plot the BPT composite objects. We show the empirical separation between AGNs and star-forming galaxies used in \citet{trouille2011}. The SALT objects with measured TBT emission lines occupy the region dominated by AGNs.} 
	\epsscale{1.}
         \end{figure} 

	 The objects that comprise our sample were selected from their IR colors to be AGNs. To test the success of our selection criteria, we sought to examine whether excitation of the narrow emission lines in our objects was driven by AGN activity or star-formation. We can compare the strong optical emission lines present in our spectra in order to explore the source of the excitation. We fit emission lines in our observed spectra with a single Gaussian. For the narrow emission lines, we forced the width of each fitted line to be the same. To estimate uncertainties on the line fluxes, we used a Monte Carlo approach. For each object we generated 500 artificial spectra by perturbing the flux at each wavelength of  the true spectrum by a random amount consistent with the 1$\sigma$ error spectrum. For each artificial spectrum, we measured the line fluxes using the same procedure as was done on the true spectrum. The standard deviation of the the distribution of line fluxes measured from the artificial spectrum was used as the error on the true line flux measurement. For each object, if a line was detected with less than 3$\sigma$ significance we used a measure of 3$\sigma$ as the maximum flux for the line, and report an upper limit. 
	
	 For nearby galaxies the optical emission-line ratios of [OIII]$\lambda$5007 / H$\beta$ as well as [NII]$\lambda$6584 / H$\alpha$ are commonly compared to separate Type II AGNs from star-forming galaxies \citep[the ``BPT Diagram''][]{baldwin1981}. Collisionally excited emission lines [OIII]$\lambda$5007 and [NII]$\lambda$6584 are much stronger in AGNs, which produce a much harder ionizing spectrum, relative to the recombination lines H$\alpha$ or H$\beta$. In the BPT Diagram, the position of an object falls into one of three commonly defined areas based on the emission-line ratios of [OIII]$\lambda$5007 / H$\beta$ as well as [NII]$\lambda$6584 / H$\alpha$: an AGN region, a star-forming locus, and a region where ``composite'' objects with evidence for both AGN activity and star-formation are found. The usage of these emission lines is limited, however, to redshifts where the emission lines are still present in the wavelength range covered by the spectrograph. The SALT RSS set-up that was used for this study spans an observed wavelength range to 6100 \AA, which excludes [NII]$\lambda$6584 and H$\alpha$, and we can only observe [OIII]$\lambda$5007 to $z < 0.22$. 

	 There are three objects in our assembled sample (J1151-0046, J1007+0146, J1320-0102) with  $z < 0.22$, and for these objects, we observe [OIII]$\lambda$5007 and H$\beta$ with high S/N. For these sources, log([OIII]$\lambda$5007 / H$\beta$)$ \geq 0.9$, a ratio which is strongly indicative of AGN activity. For the two objects with SDSS spectra, we calculated the [OIII]/H$\beta$ ratio as well. For J0325-0032, log([OIII]$\lambda$5007 / H$\beta$)$ = 1.1$ and log([NII]$\lambda$6584 / H$\alpha$)$ = -0.11$, which puts this object firmly in the AGN regime on the BPT diagram \citep{kauffmann2003b,kewley2006}. For J0924+0027, log([OIII]$\lambda$5007 / H$\beta$)$ = 0.7$, and as $z = 0.660$, the SDSS spectrum does not cover [NII] or H$\alpha$. Additionally, we can use the [OIII] luminosity for those objects in our sample with [OIII] detections to confirm the results for the luminosity regime spanned by our sample. While the fixed primary mirror of SALT prevents us from measuring the luminosity for the [OIII] lines in the spectra where the line appears (due to our inability to perform absolute flux calibration), we measured the [OIII] luminosity for the two objects with SDSS spectra. For J0924+0027, we measure $\log(L_{\mathrm{[OIII]}} / \mathrm{erg \, s}^{-1}) = 42.79$, while for J0325-0032, we measure $\log(L_{\mathrm{[OIII]}} / \mathrm{erg \, s}^{-1}) = 42.63$. Both of these values are well above the limit of $\log(L_{\mathrm{[OIII]}} / \mathrm{erg \, s}^{-1}) > 41.88$ used by \citet{reyes2008} to select Type II quasars from SDSS. 
	 	
	For objects with $z > 0.22$, we cannot use [OIII]$\lambda$5007 / H$\beta$ to test the ionization source. Instead, we used a different emission line diagnostic, [NeIII]$\lambda$3869 / [OII]$\lambda\lambda$3726+3729. This line ratio is used in the ``TBT'' Diagram \citep{trouille2011}, a plot that compares [NeIII]/[OII] to the rest-frame $g-z$ color (which we will refer to as $^{0.0}(g-z)$) for galaxies at intermediate redshifts, and can be used to separate AGNs and star-forming galaxies with a high degree of confidence. While [NeIII]$\lambda$3869 is weaker than the strong [OIII]$\lambda$5007 emission line, it is also indicative of highly ionized gas excited by an AGN, and separates AGNs from metal-rich star-forming galaxies. The rest-frame $g-z$ color helps to separate AGNs from metal-poor star-forming galaxies, as they tend to be dominated by their bulge components, and are significantly redder. \citet{trouille2011} demonstrate the success of the TBT diagram in recovering both SDSS BPT-selected AGNs (98.7\% of the objects in their sample) as well as BPT-selected star-forming galaxies (97\%). 
	
	Our usage of the TBT Diagram is limited to $z < 0.59$ where [NeIII]$\lambda$3869 appears on the RSS chip in our observational set-up. We measured the line flux ratios between the [NeIII]$\lambda$3869 and [OII]$\lambda\lambda$3726+3729 emission features, where we fit both lines simultaneously. For six objects, either [NeIII]$\lambda$3869 or [OII]$\lambda\lambda$3726+3729 fell on a gap between the detector chips or on a bright sky feature, and as a result, we cannot report a line flux ratio. For J0325-0032, we can also compare the line flux ratio measured from our SALT spectrum against the ratio measured from the SDSS spectrum. For this object, we calculate log([NeIII]/[OII])$ = -0.36 \pm 0.02$ from the SALT spectrum, and log([NeIII]/[OII])$ = -0.39 \pm 0.02$ from the SDSS spectrum, consistent within the given uncertainties.
	
	To determine $^{0.0}(g-z)$ we start with the SDSS $ugriz$ photometry and follow the methodology from \citet{trouille2011}, using kcorrect v4\_2 \citep{blanton2007} to estimate the rest-frame $g$ and $z$ magnitudes. Uncertainties on the rest-frame colors are estimated by creating 500 artificial sets of photometry for each object, where each artificial magnitude was created by randomly altering the observed SDSS photometry by an amount consistent with the uncertainties. We then ran the kcorrect software on each set of artificial magnitudes, and report the standard deviation on the distribution of resulting $^{0.0}(g-z)$ values as the uncertainty in the true value. We report both the [NeIII]$\lambda$3869 / [OII]$\lambda\lambda$3726+3729 line ratios as well as the $^{0.0}(g-z)$ values for each of the objects in our sample in Table \ref{tab:samplespec}. 
		
	We plot those targets with [NeIII]$\lambda$3869 / [OII]$\lambda\lambda$3726+3729 ratios on the TBT diagram in Figure \ref{fig:tbtdiagram}. This figure also includes contours for a large sample of SDSS emission-line galaxies at $0.02 < z < 0.35$ from \citet{trouille2011}, with emission-line fluxes measured by MPA-JHU \citep{aihara2011}. Contours for BPT-selected AGNs are shown in green, BPT star-forming galaxies are shown in blue, and BPT composite objects are shown in grey. The objects in our sample are plotted with red and blue circles depending on their selection group, and we plot those objects with upper limits on the flux ratio with an arrow\footnote{In order to focus on the bulk of the population, one object with an upper limit of [NeIII] / [OII] $< 5.26$, J1255-0010, does not appear on the plot.}. Our sources are mostly found clustered to the top of the figure, located in the region of the diagram dominated by BPT AGNs, and are distinctly separate from star-forming galaxies.
	
	For those objects for which we can use emission lines (either optical or UV) to estimate the source of the ionizing radiation, there is strong evidence that the WISE selection criteria is targeting AGNs. However, there does exist a sample of objects without optical evidence for AGN activity. There are four objects with upper limits on the ratio of [NeIII]/[OII]: J1136+0152, J1255-0010, J1320-0102, and J1609-0004. As discussed above, J1320-0102 is at a low redshift and we measure the flux ratio of log([OIII] / H$\beta$) $= 0.9$, which is indicative of an AGN. For the other three objects, if they are indeed star-forming galaxies rather than AGNs, this implies a contamination fraction in our sample of $3 / 23$ ($\sim 13\%$) (We only compare these three objects to those galaxies where we can measure the [OIII]/H$\beta$, [NeIII] / [OII], or UV emission lines). This contamination fraction agrees with the fraction of non-active galaxies observed in the \emph{Spitzer}-selected AGN sample presented in \citet{lacy2013} (22\%). If we include the 5 objects in our sample without spectroscopic redshifts, assuming that they are star-forming galaxies, the contamination fraction would have an upper limit of $\sim30$\% (8/28). In the next section, we discuss the use of SED modeling to further explore the AGN activity in our sample.
	
\section{SED Modeling}
\label{sec:sedmodeling}

	\begin{figure*}[htbp]
	\epsscale{1.0}          
 	\plotone{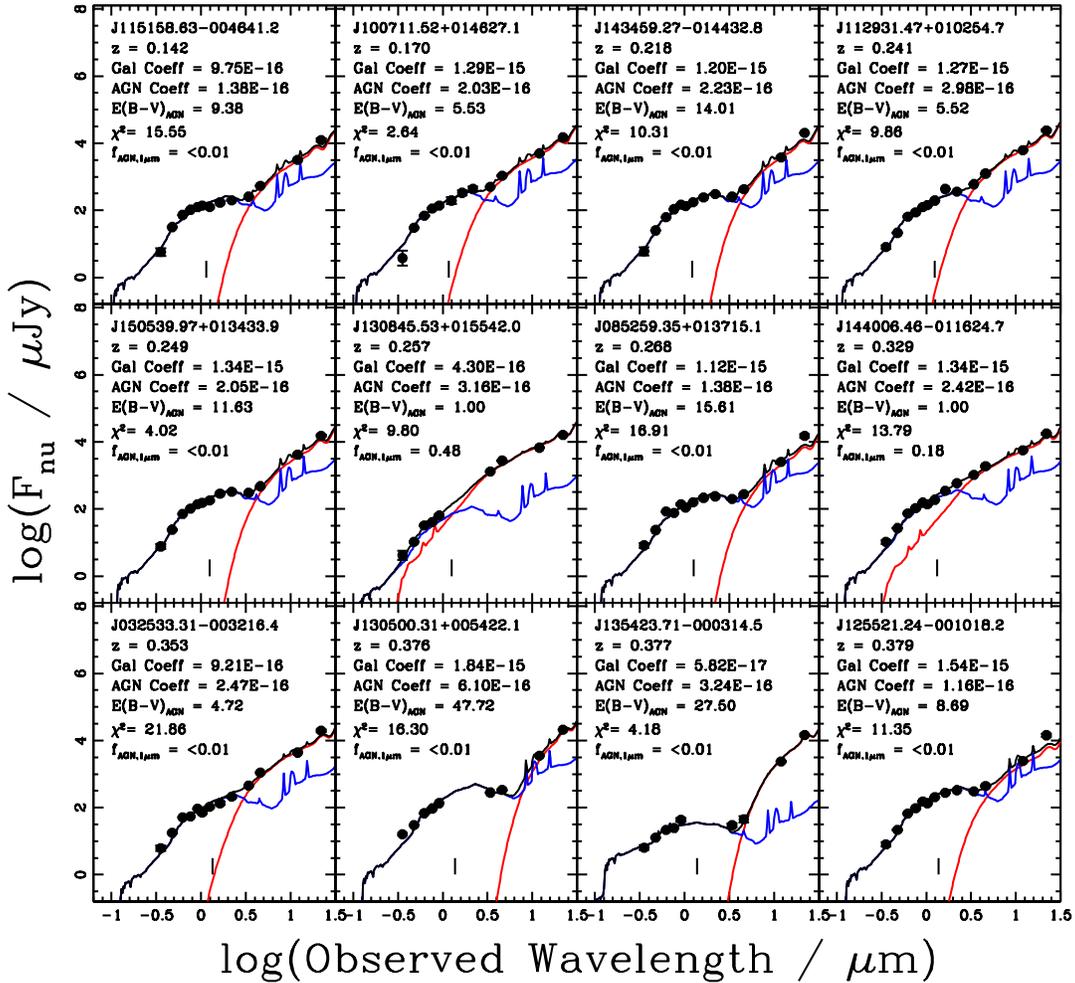}
 	\caption{\label{fig:sedmodelsone} Best-fitting SED models for the objects in our sample with spectroscopic redshifts. We fit each object in our sample with a galaxy model (blue) combined with a dust-reddened AGN model (red), using templates from \citet{assef2010}. The fit-parameters and best-fit $\chi^2$ are given in each panel for that specific fit. We also provide values for $f_{AGN, 1 \mu m}$, the fraction of the total emission at rest-frame 1$\mu$m from AGN emission derived from the fit. We show rest-frame 1$\mu$m for each of the objects with a short vertical line below the fit. Finally, we indicate photometry that was not used for the fit with open circles.} 
	\epsscale{1.}
         \end{figure*} 

	\begin{figure*}[htbp]
	\epsscale{1.0}          
 	\plotone{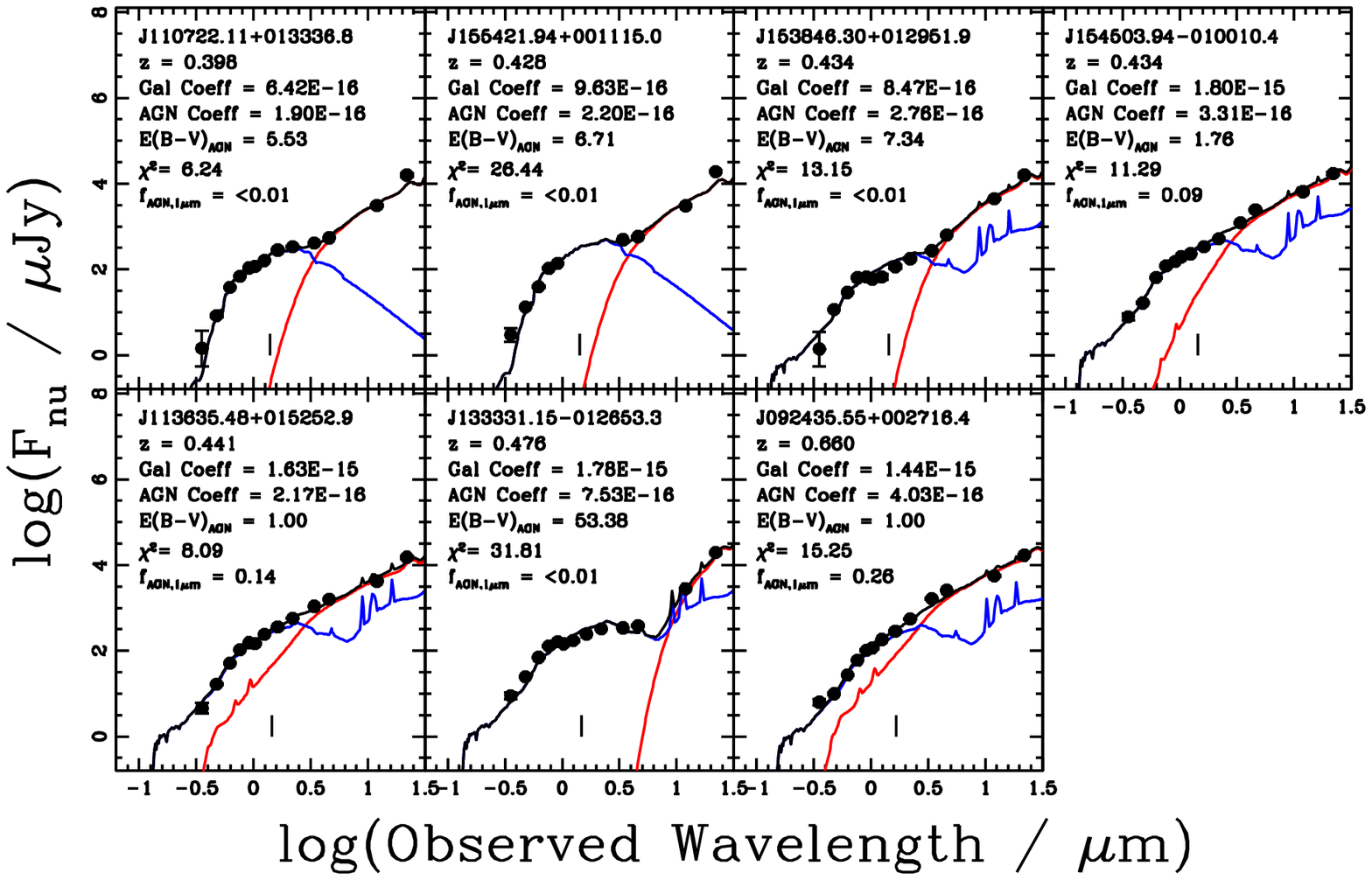}
 	\caption{\label{fig:sedmodelstwo} Continued from Figure \ref{fig:sedmodelsone}.} 
	\epsscale{1.}
         \end{figure*} 

	\begin{figure*}[htbp]
	\epsscale{1.0}          
 	\plotone{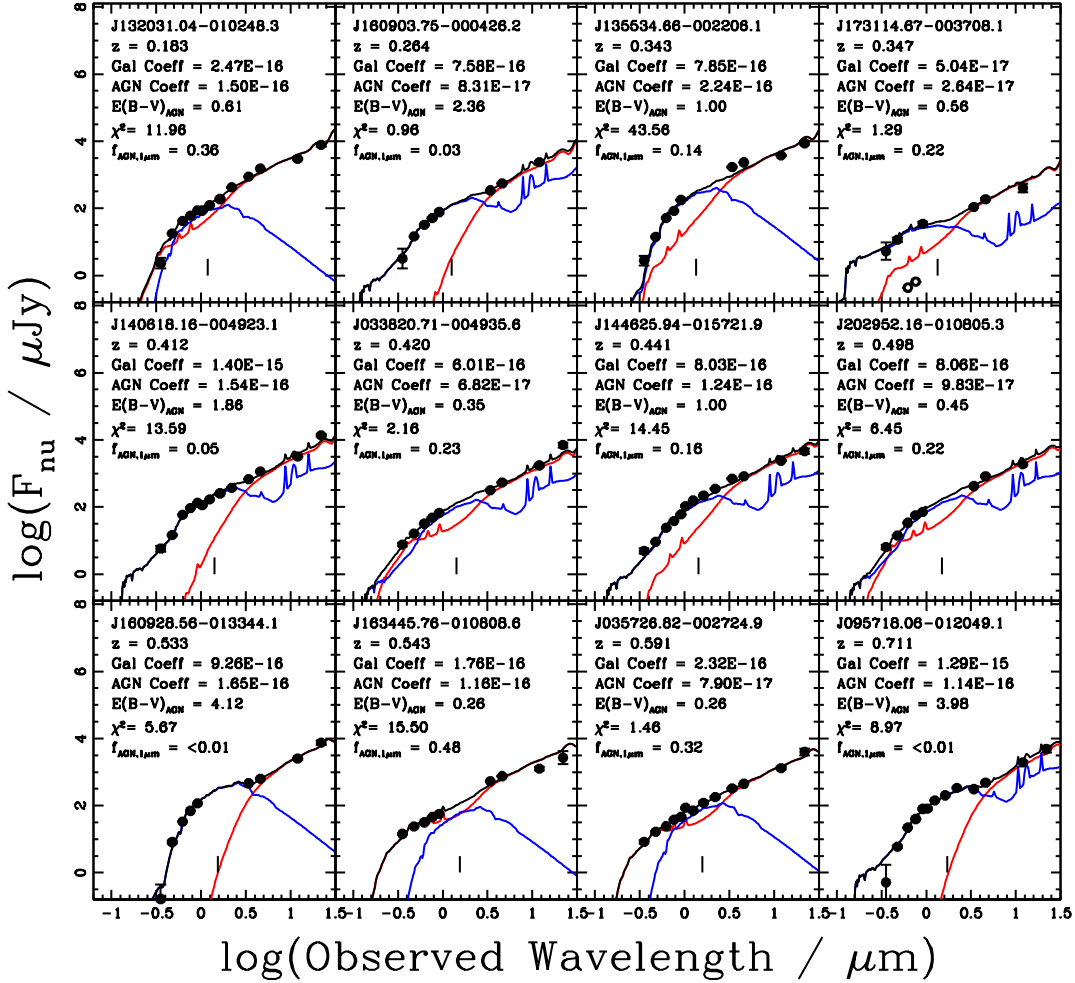}
 	\caption{\label{fig:sedmodelsthree} Best-fitting SED models for the objects in our sample with spectroscopic redshifts, for Group 2. The lines and points are the same as in Figure \ref{fig:sedmodelsone}.}  
	\epsscale{1.}
         \end{figure*} 

	\begin{figure*}[htbp]
	\epsscale{1.0}          
 	\plotone{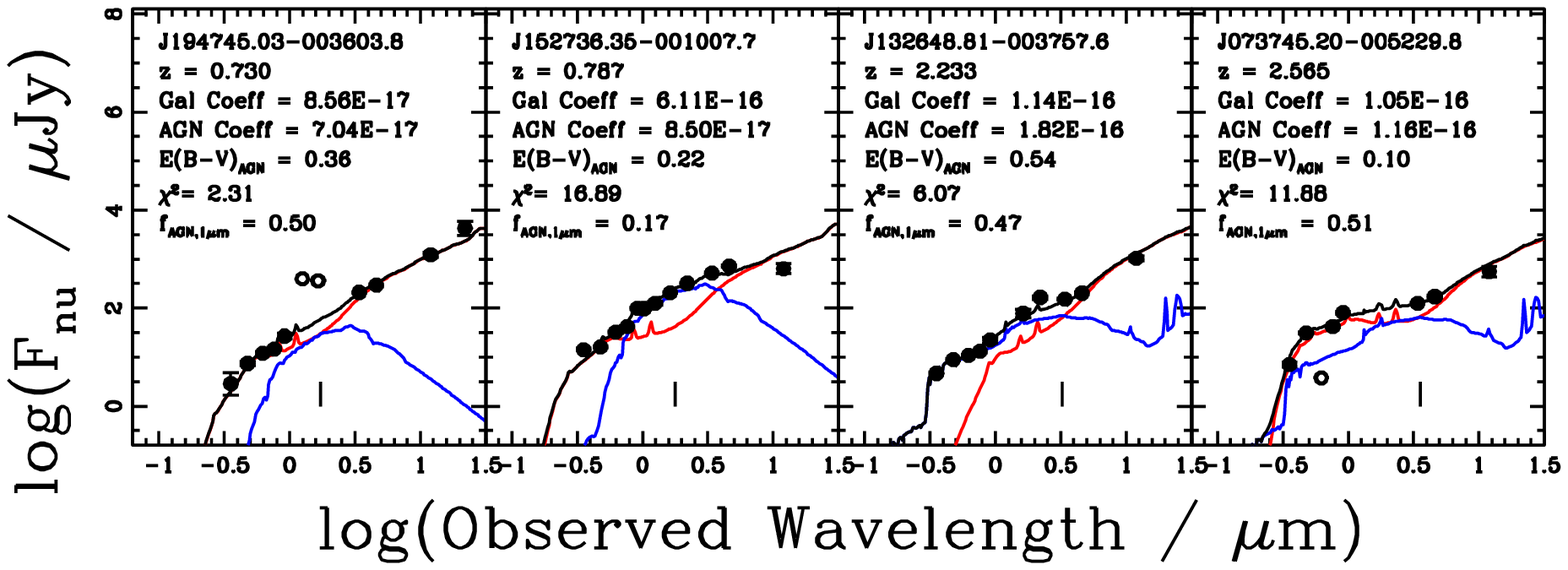}
 	\caption{\label{fig:sedmodelsfour} Continued from Figure \ref{fig:sedmodelsthree}.} 
	\epsscale{1.}
         \end{figure*} 

Existing optical through infrared photometry for the objects comprising our sample allows us to explore their spectral energy distributions (SEDs). SED decomposition is a powerful tool for gaining physical insight into the AGN and host galaxy properties of our sample, and it can also be used to estimate the redshift for those objects where we do not observe emission features in the spectrum. In order to best model the SEDs, we supplement the SDSS and WISE photometry described in Section \ref{sec:sample} with near-IR $Y$, $J$, $H$, and $K$ photometry from the UKIRT Infrared Deep Sky Survey \citep[UKIDSS,][]{lawrence2007} Large Area Survey (LAS). We use the extended source aperture-corrected magnitudes with a $5\secpoint7$ aperture measured with $>3\sigma$ significance. We have UKIDSS data for 23 objects. For an additional three objects where UKIDSS data was not available, we also used near-IR $J$, $H$, and $K$ photometry from the 2 Micron All-Sky Survey \citep[2MASS,][]{skrutskie2006} Point Source Catalogue where an object was detected in an individual filter with $>3\sigma$ significance. As this SED modeling is only designed to explore the relationship between the host galaxy and quasar emission for each object in our sample, we use the catalogue uncertainties for estimating the best-fits, but, as we are not accounting for all of the observational errors, we follow \citet{chung2014} and assign a conservative minimum photometric error of 0.05 mags on each measurement, which is comparable to or smaller than the photometric calibration uncertainties for each survey \citep{hewett2006, skrutskie2006, ahn2012, cutri2013b}. 

\begin{deluxetable*}{lclccclcr}
\tabletypesize{\scriptsize}
\tablecaption{Best-Fit SED parameters \label{tab:sedparameters}}
\tablewidth{0pt}
\tablehead{
\colhead{Object} & \colhead{Group} & \colhead{Template\tablenotemark{a}} & \colhead{Gal. Coeff\tablenotemark{a}} & \colhead{AGN Coeff\tablenotemark{a}} & \colhead{$E(B-V)_{\mathrm{AGN}}$\tablenotemark{a}} & \colhead{$\chi^2_{red}$\tablenotemark{a}} & \colhead{$f_{AGN, 1 \mu m}$\tablenotemark{b}}
}
\startdata

J032533.31-003216.4 & 1 & Sbc & 9.21e-16 & 2.47e-16 & 4.72 & 21.86 & $<$0.01 \\
J033820.71-004935.6 & 2 & Sbc & 6.01e-16 & 6.82e-17 & 0.35 & 2.16 & 0.23 \\ 
J035726.82-002724.9 & 2 & E & 2.32e-16 & 7.90e-17 & 0.26 & 1.46 & 0.32 \\ 
J073745.20-005229.8 & 2 & Im & 1.05e-16 & 1.16e-16 & 0.10 & 11.88 & 0.51 \\ 
J085259.35+013715.1 & 1 & Sbc & 1.12e-15 & 1.38e-16 & 15.61 & 16.91 & $<$0.01 \\ 
J092435.55+002716.4 & 1 & Sbc & 1.44e-15 & 4.03e-16 & 1.00\tablenotemark{c} & 15.25 & 0.26 \\ 
J095718.06-012049.1 & 2 & Sbc & 1.29e-15 & 1.14e-16 & 3.98 & 8.97 & $<$0.01 \\ 
J100711.52+014627.1 & 1 & Sbc & 1.29e-15 & 2.03e-16 & 5.53 & 2.64 & $<$0.01 \\ 
J110722.11+013336.8 & 1 & E & 6.42e-16 & 1.90e-16 & 5.53 & 6.24 & $<$0.01 \\ 
J112931.47+010254.7 & 1 & Sbc & 1.27e-15 & 2.98e-16 & 5.52 & 9.86 & $<$0.01  \\
J113635.48+015252.9 & 1 & Sbc & 1.63e-15 & 2.17e-16 & 1.00\tablenotemark{c} & 8.09 & 0.14 \\ 
J115158.63-004641.2 & 1 & Sbc & 9.75e-16 & 1.38e-16 & 9.38 & 15.55 & $<$0.01 \\ 
J125521.24-001018.2 & 1 & Sbc & 1.54e-15 & 1.16e-16 & 8.69 & 11.35 & $<$0.01 \\ 
J130500.31+005422.1 & 1 & Sbc & 1.84e-15 & 6.10e-16 & 47.72 & 16.30 & $<$0.01 \\ 
J130845.53+015542.0 & 1 & Sbc & 4.30e-16 & 3.16e-16 & 1.00\tablenotemark{c} & 9.80 & 0.48 \\ 
J132031.04-010248.3 & 2 & E & 2.47e-16 & 1.50e-16 & 0.61 & 11.96 & 0.36 \\ 
J132648.81-003757.6 & 2 & Im & 1.14e-16 & 1.82e-16 & 0.54 & 6.07 & 0.47 \\ 
J133331.15-012653.3 & 1 & Sbc & 1.78e-15 & 7.53e-16 & 53.38 & 31.81 & $<$0.01 \\ 
J135423.71-000314.5 & 1 & Im & 5.82e-17 & 3.24e-16 & 27.50 & 4.18 & $<$0.01 \\ 
J135534.66-002206.1 & 2 & E & 7.85e-16 & 2.24e-16 & 1.00\tablenotemark{c} & 43.56 & 0.14 \\ 
J140618.16-004923.1 & 2 & Sbc & 1.40e-15 & 1.54e-16 & 1.86 & 13.59 & 0.05 \\ 
J143459.27-014432.8 & 2 & Sbc & 1.20e-15 & 2.23e-16 & 14.01 & 10.31 & $<$0.01 \\ 
J144006.46-011624.7 & 1 & Sbc & 1.34e-15 & 2.42e-16 & 1.00\tablenotemark{c} & 13.79 & 0.18 \\ 
J144625.94-015721.9 & 2 & Sbc & 8.03e-16 & 1.24e-16 & 1.00\tablenotemark{c} & 14.45 & 0.16 \\ 
J150539.97+013433.9 & 1 & Sbc & 1.34e-15 & 2.05e-16 & 11.63 & 4.02 & $<$0.01 \\ 
J152736.35-001007.7 & 2 & E & 6.11e-16 & 8.50e-17 & 0.22 & 16.89 & 0.17 \\ 
J153846.30+012951.9 & 1 & Sbc & 8.47e-16 & 2.76e-16 & 7.34 & 13.15 & $<$0.01 \\ 
J154503.94-010010.4 & 1 & Sbc & 1.80e-15 & 3.31e-16 & 1.76 & 11.29 & 0.09 \\ 
J155421.94+001115.0 & 1 & E & 9.63e-16 & 2.20e-16 & 6.71 & 26.44 & $<$0.01 \\ 
J160903.75-000426.2 & 2 & Sbc & 7.58e-16 & 8.31e-17 & 2.36 & 0.96 & 0.03 \\ 
J160928.56-013344.1 & 2 & E & 9.26e-16 & 1.65e-16 & 4.12 & 5.67 & $<$0.01 \\ 
J163445.76-010808.6 & 2 & E & 1.76e-16 & 1.16e-16 & 0.26 & 15.50 & 0.48 \\ 
J173114.67-003708.1 & 2 & Im & 5.04e-17 & 2.64e-17 & 0.56 & 1.29 & 0.22 \\ 
J194745.03-003603.8 & 2 & E & 8.56e-17 & 7.04e-17 & 0.36 & 2.31 & 0.50 \\ 
J202952.16-010805.3 & 2 & Sbc & 8.06e-16 & 9.83e-17 & 0.45 & 6.45 & 0.22 
\enddata
\tablenotetext{a}{Parameters from the SED fits using the templates from \citet{assef2010}. Each fit was done using one galaxy template, along with the dust obscured AGN template with obscuration parameterized by $E(B-V)_{\mathrm{AGN}}$. We also provide the reduced $\chi^2$ value for each fit.}
\tablenotetext{b}{Fraction of the total emission at rest-frame 1$\mu$m that is estimated to arise from AGN emission.}
\tablenotetext{c}{For these objects, the fits were restricted to $E(B-V)_{\mathrm{AGN}} \ge 1.0$ due to the objects being resolved in optical SDSS imaging, and most likely unobscured AGNs.}
\end{deluxetable*}

For the fitting, we follow the methodology we model each galaxy as a linear combination of a single galaxy template and a reddened AGN template (note that this is different from the methodology discussed in \citet{assef2010}, where each galaxy was modeled as a linear combination of all three galaxy templates plus an AGN template). We use the empirically-derived models from \citet{assef2010}, which consist of three galaxy templates (a spiral ``Sbc'', elliptical ``E'', and irregular ``Im'' galaxy), as well as an unobscured AGN template. In order to simulate the effects of AGN obscuration, we follow \citet{assef2010} and apply an extinction model to the AGN template that consists of a Small Magellanic Cloud (SMC) like extinction curve at $\lambda < 3300$\AA\, \citep{gordon1998} and a Galactic extinction curve at longer wavelengths \citep{cardelli1989}, with $R_{V} = 3.1$ for both, and in this paper we parameterize the extinction using $E(B-V)$. Because we are attempting to fit an unobscured AGN template to objects with varying amounts of dust obscuration, we follow \citet{mainieri2011} and \cite{hainline2012} and employ an additional prior of $E(B-V)_{\mathrm{AGN}} > 1.0$ for those objects with evidence indicating that they are an obscured AGN. We use this prior on objects with an SDSS TYPE flag of ``3'', indicating that the galaxy is resolved in optical imaging. For these objects, we assume that the optical portion of the SED is dominated by stellar emission rather than that of the central AGN. Setting a lower bound on the best-fit extinction value was only required for fitting 6 objects out of the total 35 that we modeled in this way, and the fits were not made significantly worse by using this prior. For each object, we calculated the monochromatic flux densities from the observed photometry\footnote{For the WISE photometry, we used the conversions provided in Section IV.4.h.i of the WISE All-Sky Data Release Explanatory Supplement.}, and compared these values to the flux densities from the best-fit templates passed through the SDSS, UKIDSS, 2MASS, and WISE bandpass filters, and used $\chi^2$ minimization to find the best fits for each object. The output parameters are then two coefficients for the linear combination (a galaxy and an AGN coefficient), and a value for $E(B-V)_{\mathrm{AGN}}$.   The results from our fitting are shown in Table \ref{tab:sedparameters} and Figures \ref{fig:sedmodelsone}, \ref{fig:sedmodelstwo}, \ref{fig:sedmodelsthree}, and \ref{fig:sedmodelsfour}. 

As can be seen from the best-fit $\chi^2$ values, the majority of the objects in our sample are well fit by the \citet{assef2010} templates. J1355-0022 has $\chi^2_{red} > 40$, due primarily to the \citeauthor{assef2010} AGN template being redder than the observed W3$-$W4 color. We observe a large variety in the optical SED shapes for the objects in our sample, and all three galaxy templates are used in our best fit models. For each object, the best-fit AGN coefficients are above zero, indicating that galaxy light alone is unlikely to be the source of the infrared emission seen in the SEDs. For the three objects described in the previous section that do not show evidence in their optical spectra for an AGN, we do recover non-zero AGN coefficients, but these SEDs would also perhaps be fit using a more extreme star-forming template than the one derived in \citet{assef2010}. Indeed, the SEDs of both M82 and Arp220 taken from the SWIRE template library \citep{polletta2007} have stronger mid-IR emission from dust heated by star formation, which fits the WISE photometry for the objects in our sample with the highest values for $E(B-V)_{\mathrm{AGN}}$. For each object, we also calculate $f_{AGN, 1 \mu m}$, which is defined as the fraction of the total emission in the best-fit model at rest-frame 1 $\mu$m that is contributed by the AGN template \citep{hainline2012}. We list these values in Table \ref{tab:sedparameters}. There are 16 objects with $f_{AGN, 1 \mu m} < 0.01$, which we note in the table. To check the validity of our SED fits, we examined the far-infrared luminosities and star-formation rates (SFRs) implied by the best-fit galaxy coefficients. The \citet{assef2010} templates only extend to a rest-frame wavelength of $\sim 30 \mu$m, and so we used \citet{charyelbaz2001} far-infrared star-forming galaxy templates to predict the far-infrared luminosity from star-formation for our sample. We passed the observed best-fit galaxy template (ignoring the AGN component) through the \textit{Spitzer} MIPS 24 $\mu$m passband, and then used this flux to constrain the \citet{charyelbaz2001} templates, which were integrated to produce the far-infrared luminosity from star-formation for each object. We then used the conversion from \citet{kennicutt1998} to calculate the SFRs. We measure an average (median) SFR for the sample (without the two high-redshift objects) of $16\, (6)$ M$_{\sun}$ yr$^{-1}$, in agreement with expectations for massive star-forming galaxies at the redshift range probed by our sample. We note that this method is highly uncertain, especially without longer-wavelength observations to better constrain the far-infrared templates.

	\begin{figure}[htbp]
	\epsscale{1.2}          
 	\plotone{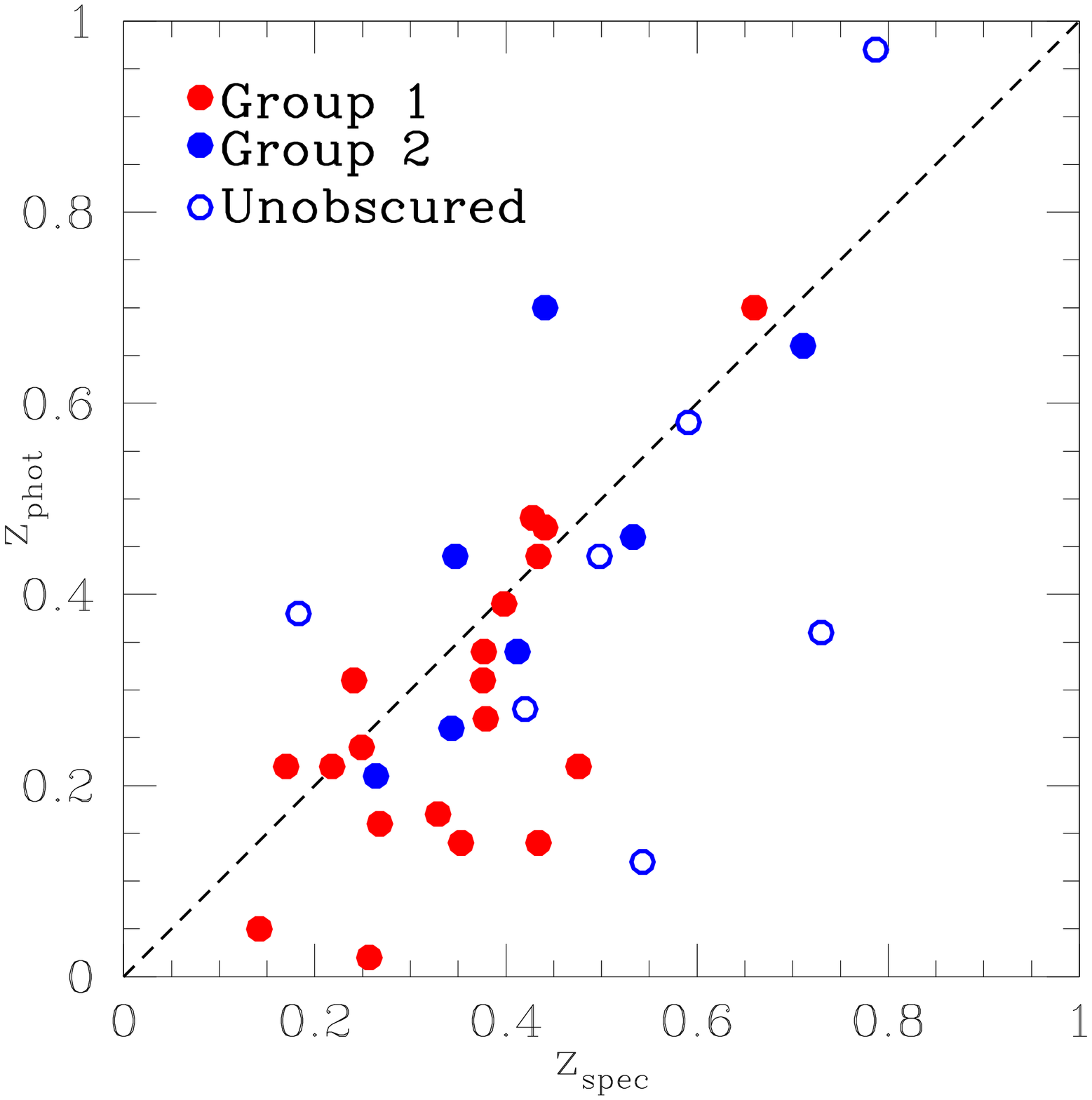}
 	\caption{
	\label{fig:specz_vs_photz} Spectroscopic redshift plotted against derived photometric redshift for the objects in our sample. We plot our points from Group 1 in red and Group 2 in blue, and we plot those objects with SED evidence for being unobscured with an open circle. While the majority of the objects have photometric redshifts that are similar, if slightly lower than their spectroscopic redshifts, there is a sample of AGN dominated sources which have significantly lower photometric redshifts, due to the difficulty in fitting an SED to a featureless AGN power-law SED, as seen in Figures \ref{fig:sedmodelsone} - \ref{fig:sedmodelsfour}.} 
	\epsscale{1.}
         \end{figure} 

In order to estimate redshifts for those objects in our sample without detected emission lines, we ran the SED analysis as described above on each of the objects in our sample but allowed redshift to be a free parameter. Estimating redshifts from AGN photometry is difficult as the power-law shape of an AGN SED with minimal extinction creates a degeneracy between color and redshift which requires long wavelength photometric coverage to break \citep{brodwin2006,salvato2009,salvato2011}. We examined the photometric redshifts recovered for those objects with spectroscopic redshifts, as well as those with existing SDSS photometric redshifts \citep{oyaizu2008} (photo-z's were estimated for each object that was resolved in SDSS imaging, and these values largely agree with the spectroscopic redshifts for these objects). For our own photometric redshift calculation, we used the $\chi^2$ values to explore the 2$\sigma$ confidence intervals and find the best physical fit to the data for each object, incorporating the $E(B-V)_{\mathrm{AGN}} > 1.0$ prior discussed above for objects that were resolved in SDSS imaging. In a few cases, we rejected best-fits at large redshifts ($z > 2$) both due to the low likelihood of a galaxy with $20 < g < 22$ to be at higher redshift (See our analysis of the AGNs in this optical photometric range in the Bo\"{o}tes field in Section \ref{sec:opticalspectra}) and the low likelihood of a quasar to be this luminous at such high redshift, based on the observed quasar luminosity function \citep{hopkins2007}. The final photometric redshifts estimated from this procedure are given in Table \ref{tab:samplespec}. In Figure \ref{fig:specz_vs_photz} we plot the spectroscopic redshifts we derived from our optical spectra against the photometric redshifts we derived from this process for objects with $z < 1.0$. Overall, there is excellent agreement between the spectroscopic and photometric redshifts for the majority of the objects in our sample, although our recovered photometric redshifts are often low in comparison to the corresponding spectroscopic redshifts. The two most discrepant objects, J1634-0108 and J1947-0036 have SEDs that are dominated by AGN emission ($f_{AGN, 1 \mu m} \sim 0.5$, and $E(B-V)_{\mathrm{AGN}} < 0.4$).

\begin{figure*}[htbp]
	\epsscale{1.0}          
 	\plotone{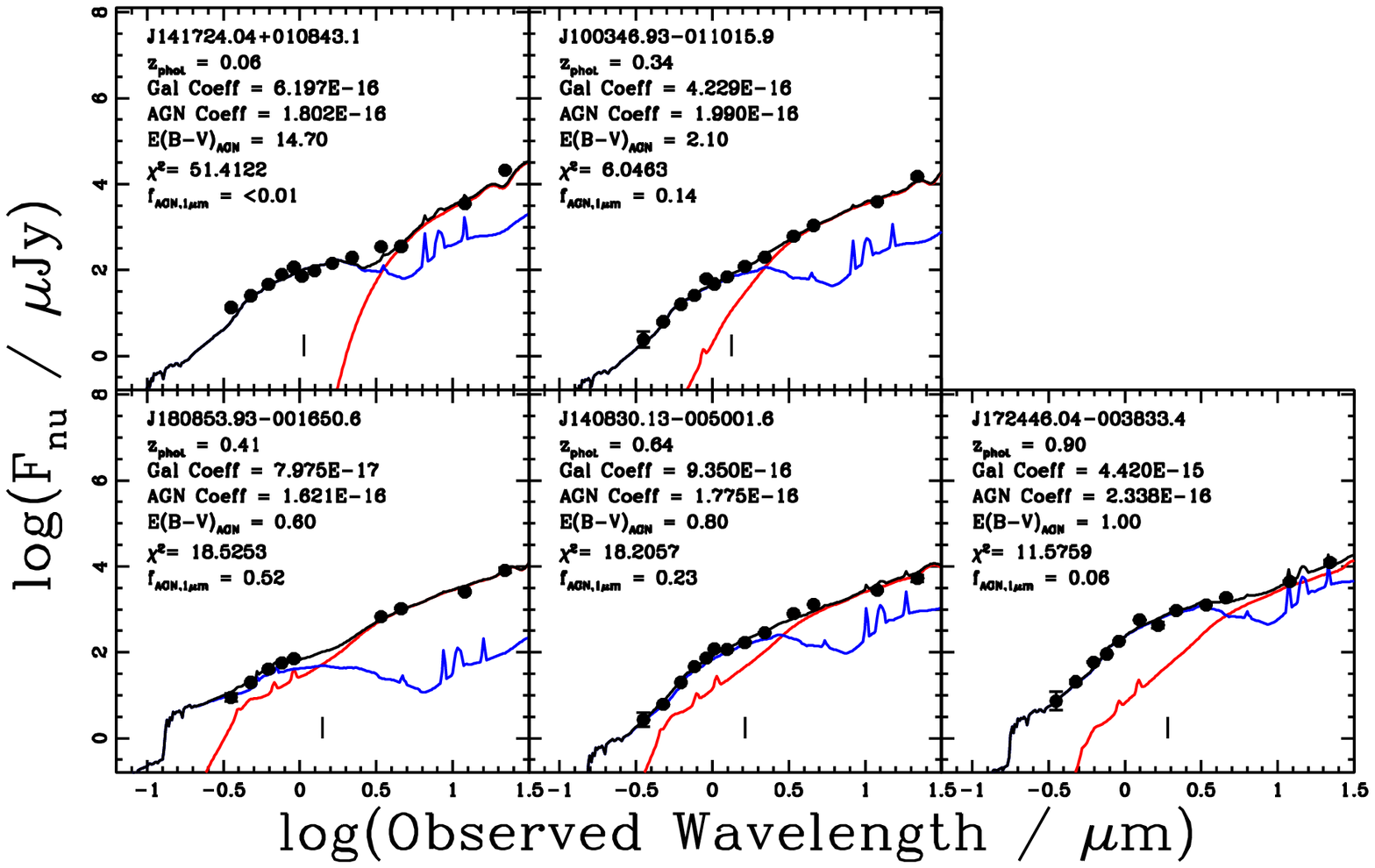}
 	\caption{\label{fig:sedmodelsphotz} Best-fitting SED models for the objects in our sample without spectroscopic redshifts. The lines and points are the same as in Figure \ref{fig:sedmodelsone}. The photometric redshifts are given in each plot. The top two objects are from Group 1, while the bottom three objects are from Group 2.} 
	\epsscale{1.}
         \end{figure*}

With these uncertainties in mind, we estimated the photometric redshift for the five objects in our sample without spectroscopic redshifts seen in Figure \ref{fig:group3redshifts}. We plot the best-fit SEDs in Figure \ref{fig:sedmodelsphotz} and report the calculated photometric redshifts for these sources in Table \ref{tab:samplespec}. As can be seen, these objects have a range of $0.1 < z_{phot} < 0.9$, in broad agreement with the spectroscopic redshifts observed for our sample. If these photometric redshifts are roughly correct, then we calculate an average redshift of $z = 0.52$ for all of the objects in our sample, while Group 1 has an average redshift of $z = 0.36$, and Group 2 has an average redshift of $z = 0.71$ (the average redshift is $z = 0.52$ without including the two $z \sim 2$ objects), supporting our conclusions that selecting objects based on fainter W4 photometry leads to a higher redshift sample. It would be of special interest to extend this photometric redshift estimation to a larger sample of SDSS+WISE-selected AGNs in order to support quasar clustering analyses (e.g. DiPompeo et al. 2014).

\section{Individual Objects}
\label{sec:individualobjects}

While we discussed the full population of objects in Sections \ref{sec:opticalspectra} and \ref{sec:linediagnostics}, some of the individual objects that comprise the sample warrant a more in-depth discussion of their optical spectroscopic and SED properties. In this section, we will describe our highest redshift sources, as well as those sources with broad emission lines in their spectra. 

\subsection{High-Redshift Sources}
\label{sec:highredshift}

Two of the objects in our sample, J0737-0052 ($z = 2.565$) and J1326-0037 ($z = 2.233$), were identified by strong UV emission lines in their observed spectrum. As discussed in Section \ref{sec:opticalspectra}, it is not surprising to find objects in our sample at these redshifts, as the redshift distribution of objects selected to be AGNs by their WISE colors seen in many other studies commonly extends to $z \sim 2$ and above. Both of the high-redshift sources we describe have spectra with Ly$\alpha$ in emission, and both show evidence for CIV$\lambda$1548 emission, while J0737-0052 also shows strong NV$\lambda$1240 emission. These UV emission features are strongly indicative of the presence of an AGN \citep[e.g.][]{hainline2011}, and based on the 8$\mu$m luminosities calculated for these objects, they are in the quasar regime.

The UV spectrum of J0737-0052 shows additional broad absorption features on each of the strong emission lines, indicating that this object is a Broad Absorption Line (BAL) quasar \citep{foltz1990, weymann1991, reichard2003, dipompeo2012}, with absorption troughs having FWHM$ > 2000$ km s$^{-1}$. In order to explore the broad absorption in this object we examined the CIV$\lambda$1548 emission and absorption profile, and we measure FWHM$_{\mathrm{CIV}} = 2500$ km s$^{-1}$. We also fit the CIV feature as a sum of Gaussians. The best-fit was composed of one Gaussian in emission (FWHM $ = 4080 \pm 80 $ km s$^{-1}$ ) along with two Gaussians in absorption (FWHM $ = 2400 \pm 70$ km s$^{-1}$ and FWHM $ = 1050 \pm 70$ km s$^{-1}$). While a full fit of the more complex Ly$\alpha$, NV, and SiIV 1393+1402 regions is outside the scope of this current work, it should be noted that BAL quasars have been shown to have red optical to ultraviolet continua \citep{weymann1991,sprayberry1992,dipompeo2012} and the discovery of a high-redshift BAL quasar is not uncommon in other samples of optically red WISE-selected quasars selected under different criteria \citep[e.g.][]{ross2014}.

Although the spectrum of J1326-0037 does have lower S/N, we observe both the Ly$\alpha$  and CIV$\lambda$1548 emission lines. It is interesting to note that the Ly$\alpha$ emission line has two peaks, which may be due to intrinsic Ly$\alpha$ absorption. As we show in Figure \ref{fig:hi_z_figures}, the central absorption is also seen in the 2D spectrum. We fit the emission feature with two Gaussians, one in emission and the other in absorption, and found that the wavelength of the line centroids agreed with each other to within the uncertainty on the values, while the best-fit observed frame FWHM of the emission and absorption features, respectively, are FWHM$_{\mathrm{obs,emission}}$ = $2500 \pm 100$ km s$^{-1}$, and FWHM$_{\mathrm{obs,absorption}}$ = $350 \pm 90$ km s$^{-1}$. We also explored Gaussian fits under the assumption that the Ly$\alpha$ emission feature is instead double-peaked, and find that the blue peak has an observed frame FWHM$_{\mathrm{obs,blue}}$ = $1500 \pm 100$ km s$^{-1}$, while we measured FWHM$_{\mathrm{obs,red}}$ = $730 \pm 70$ km s$^{-1}$. In the rest-frame, the velocity difference between the peaks is $\Delta v_{\mathrm{Ly}\alpha} = -1570 \pm 40$ km s$^{-1}$. While the CIV emission line also shows evidence for central absorption, the low S/N of the feature does not allow for a robust fit to determine a two-Gaussian line profile. 

As seen in Figure \ref{fig:sedmodelsfour} and Table \ref{tab:sedparameters} both of the high-redshift quasars in our sample are well fit by the AGN templates. J1326-0037 is best fit by an Im galaxy template that is dominant at short wavelengths, indicating that the UV continuum observed in the optical spectrum may be stellar in origin. As would be expected from the optical spectrum, J0737-0052 has an SED that is dominated by AGN emission, with $f_{AGN, 1 \mu m} = 0.5$. Overall, both J1326-0037 and J0737-0052 represent interesting objects for potential follow-up observations.

\subsection{Unobscured Sources}
\label{sec:unobscured}

While our sample was selected to focus on obscured quasars, both the optical spectra and the SED modeling results demonstrate that there exists a significant number of unobscured objects. There are two objects with broad emission lines in our sample: J0737-0052, and J1634-0108. We discussed J0737-0052 in the previous section. For J1634-0108, which is unresolved in the SDSS imaging, we observe broad MgII$\lambda$2798, which we fit as a sum of two Gaussians representing the narrower and a broader components of the line. The comparatively narrow component of the MgII emission feature for J1634-0108 is measured to have FWHM$_{\mathrm{MgII,narrow}} = 2660 \pm 60$ km s$^{-1}$. For the broad component in J1634-0108, however, we measure FWHM$_{\mathrm{MgII,broad}} = 10,000 \pm 300$ km s$^{-1}$. Based on the presence of these broad emission lines, we classify this object as a Type I quasar. 

The remainder of our objects do not show broad emission lines in the wavelength ranged probed by our optical spectroscopy, but we can use the results from our SED modeling to demonstrate that emission at optical wavelengths likely arises from the quasar. For 8 objects, we see optical emission that is dominated by quasar light: J0338-0049, J0357-0027, J0737-0052, J1320-0102, J1527-0010, J1634-0108, J1947-0036, and J2029-0108. Two of these objects are Type I quasars from their broad emission lines. The remainder are  unresolved in SDSS imaging, and have low values for $E(B-V)_{\mathrm{AGN}}$, similar to the ``dust-reddened quasars'' discussed in \citet{urrutia2005}, \citet{banerji2012}, and \citet{glikman2013}. Most notably, all but one of these objects (J0338-0049) have $r_{\mathrm{AB}}-W2_{\mathrm{AB}} > 3.1$, the demarcation we used to select obscured objects for Group 2. Some contamination of Type I objects is to be expected based on the original optical-IR selection criteria outlined in \citet{hickox2007}, and as we selected our objects in a relatively narrow range in $g$-band magnitude, it is likely that we are not exploring the most obscured quasars, where $g > 22$ (As discussed in Section \ref{sec:sample}, there exists a significant number of these objects that would otherwise satisfy the Group 1 and Group 2 selection criteria). There are 12 objects in our sample with $r_{\mathrm{AB}}-\mathrm{W2}_{\mathrm{AB}} < 3.1$, and of those objects, only J0338-0049 has an SED that indicates that this object is unobscured. As can be seen from our SED fits, the most obscured quasars in our sample have blue $r-\mathrm{W2}$ colors, as the AGN only becomes dominant at longer wavelengths. Because the objects in Group 1 were chosen with bright W4 photometry and were limited to a specific range in $g$-band magnitude, these objects were effectively chosen using a $g-\mathrm{W4}$ color cut, which lead to the selection of the most heavily obscured sources in our sample.

\section{Comparison to other AGN Selection Criteria}
\label{sec:agnselection}

Based on our examination of the optical spectroscopic properties of our sample in Section \ref{sec:linediagnostics}, we conclude that the WISE IR selection criteria we used was largely successful at recovering obscured quasars. We compare our simple selection criteria, which was based on the W1$-$W2 color cut from \citet{stern2012}, to other criteria discussed in the literature. Recently, \citet{mateos2012} used AGNs selected using the Bright Ultrahard XMM-Newton Survey to create a WISE selection criteria that attempted to maximize the fraction of X-ray selected AGNs while minimizing contamination from star-forming galaxies. In a follow-up paper \citep{mateos2013}, the authors examine the effectiveness of their revised WISE selection criteria in selecting obscured quasars using the SDSS Type II quasar sample described in \citet{reyes2008}. Their results indicate that it is AGN luminosity, especially with respect to host galaxy luminosity, that is the driving factor in selecting AGNs by their mid-IR colors. In this section we describe where our objects lie with respect to the \citet{mateos2012} selection criteria. 

Based solely on the selection criteria described in \citet{mateos2012}, 32/40 (80\%) of the objects in our SALT spectroscopic sample fall in the \citeauthor{mateos2012} color-color wedge, as can be seen in Figure \ref{fig:wisecolorcolor}. We calculated the redshift of 28/32 (88\%) of these objects, although 4 only have one emission feature in their spectrum. If we exclude objects where the redshift precludes us discerning any AGN activity in the object (due to both the wavelength coverage of the SALT chip, or either [NeIII] or [OII] falling on a chip gap or sky feature), then the \citeauthor{mateos2012} criteria were successful at identifying 14/17 (82\%) objects with strong evidence of AGN activity based on the objects position on the TBT diagram, the BPT diagram, or the presence of high-ionization emission lines in the UV spectrum for the two highest redshift objects. The other three objects have upper limits on the [NeIII]/[OII] ratio, indicating a non-detection of [NeIII] in the spectrum. 

8 objects in our SALT sample would not be selected under the \citet{mateos2012} criteria. Two of these objects have unusual W2$-$W3 colors, J1527-0010 ($\mathrm{W2} - \mathrm{W3} = 1.80$), and J1354-0003 ($\mathrm{W2} - \mathrm{W3} = 6.24$). There is only one emission line identified in the spectrum of J1527-0010 and while J1354-0003 is in the AGN region of the TBT diagram, the position of this object is in the region also occupied by AGN/SF composite objects. 

The other 6 objects not selected by the \citet{mateos2012} wedge have red W2$-$W3 colors that put them below and to the right of the wedge. Five of these objects (J1305+0054, J0852+0137, J1554+0011, J1107+0133, and J1333-0126) are identified as AGNs based on their position on the TBT diagram, and we have not identified the spectroscopic redshift of the sixth object, J1417+0108. The presence of AGNs outside of the \citet{mateos2012} selection region is not surprising; these authors used an analysis of AGN mid-IR templates to demonstrate that heavily obscured AGNs, as well as higher redshift AGNS ($z \sim 1 - 1.5$) would occupy this position in WISE color-color space. The positions of heavily obscured AGNs in WISE color-color space was further discussed in \citet{mateos2013}, where the relative contribution of an AGN and its host to the mid-IR SED was the primary factor that determines whether or not an AGN was found in the \citeauthor{mateos2012} selection wedge. For the five objects in our sample outside the wedge with spectroscopic redshifts, the best-fit SEDs indicate that these objects have a best-fit value of $E(B-V)_{\mathrm{AGN}} > 6.0$ (three of these objects have $E(B-V)_{\mathrm{AGN}} > 16.9$), indicating that the AGN only becomes dominant in the mid-IR; this is expected based on the red W2$-$W3 colors for these objects. As mentioned in the previous section, these objects would not be chosen under the Group 2 selection criteria, as they have $r_{\mathrm{AB}}-\mathrm{W2}_{\mathrm{AB}} < 3.1$. For selecting the most highly obscured quasars, these results indicate the necessity in using an optical to infrared color cut with the W3 or W4 photometric bands. These high values for $E(B-V)$ indicate that these objects are likely obscured quasars, but it should be noted that while we do apply a simple prescription for obscuration in our SED fitting, we do not know the geometry of the obscuring medium. As we do see strong infrared emission, we know that there must be dust near the central black hole, but galaxy-scale dust can also contribute to the observed obscuration \citep[see][\& Chen et al. in prep]{merloni2014}. We cannot rule out extreme star-formation as the the origin of the rest-frame IR photometry, supported by the measurement of $f_{AGN, 1 \mu m} \leq 0.01$ for these objects and the position of the \citet{polletta2007} Arp220 template occupying a similar position on the color-color diagram at low redshift as can be seen in Figure \ref{fig:wisecolorcolor}. However this figure as well as \citet{mateos2013} demonstrate that a significant number of SDSS Type II quasars from the \citet{reyes2008} sample also fall into this region and would not be selected as IR-AGNs using the \citet{mateos2012} criteria. 

We can also use the SED modeling results to test whether our full sample of objects would be selected under the \emph{Spitzer} selection criteria outlined in \citet{lacy2004}, \citet{stern2005}, and \citet{donley2012}. We passed the best-fit combined stellar and AGN template through the \emph{Spitzer} IRAC transmission curves to estimate \emph{Spitzer} photometry. We find that all objects but one fall into the \citet{lacy2004} selection box. The lone outlier, J1354-0003, has an SED with a very strong AGN power-law as compared to the stellar population, as seen in Figure \ref{fig:sedmodelsone}. \citet{donley2012} revised the \citet{lacy2004} criteria in order to account for contamination to the IR from star-forming galaxies, and we find that only three of our targets would lie outside of the \citet{donley2012} selection box. These objects, J1333-0126, J1305+0054, J0852+0137, are objects that would also not be selected under the \citet{mateos2012} criteria and have near-IR emission dominated by host galaxy light, with our SED modeling showing that an AGN contribution is only prominent at longer wavelengths. The bulk of the targets that comprise our sample, then, compare very well to the \emph{Spitzer}-selected objects described in \citet{lacy2013}. 

\section{Conclusions}
\label{sec:conclusions}

The all-sky coverage of WISE has made it possible to find large numbers of obscured quasars based solely on their observed infrared colors. We have observed a sample of 40 objects with with WISE colors indicative of AGN activity in two groups. Group 1 objects were initially selected to have $\mathrm{W1}- \mathrm{W2} > 0.7$ and $7 \ge \mathrm{W4} \ge 6.5$, while Group 2 objects were initially selected to have $\mathrm{W1}- \mathrm{W2} > 0.8$, $\mathrm{W4} \ge 7.0$, and $r_{\mathrm{AB}}-\mathrm{W2}_{\mathrm{AB}} > 3.1$. Both groups were selected to have $20 < g < 22$. We used optical spectroscopy from SALT RSS to uncover the redshift distribution and ionization properties of obscured quasars. We use the ratio [OIII] / H$\beta$, and [NeIII] / [OII] to determine that the majority of our objects in our sample are AGNs: of those objects with redshifts that allow us to use emission lines to identify the ionization source, only $\sim 13\%$ of the objects in our sample do not have evidence in their observed spectra of being an AGN. For the AGNs, we measure [OIII] and IR luminosities that are comparable to SDSS-selected quasars. The objects in our sample have an average redshift of $z \sim 0.5$, and the majority have $z < 1.0$, which is most likely a result of our optical photometric cut. Our sample also includes two interesting high-redshift sources, a BAL quasar at $z = 2.565$, and a quasar with Ly$\alpha$ and CIV$\lambda$1549 emission at $z = 2.233$. 

We also use a simple two-component SED analysis to explore the relationship between AGN and star-formation as a function of the position of the objects on the WISE W1$-$W2 vs. W3$-$W4 color-color diagram. The majority of the objects in our sample have strong AGN emission in the mid-IR. We also compared our selection criteria to that of \citet{mateos2012}. While their criteria are successful at recovering the AGNs in our sample, there are five objects outside of their selection wedge at red W3$-$W4 colors with optical line ratios indicative of AGN activity. SED modeling of these objects demonstrates that these are the most highly obscured quasars in our sample, in line with the conclusions made in \citet{mateos2012}. These results indicate the selecting objects based on red optical to W3 or W4 colors is required to select objects with the highest levels of obscuration. 

The results from observing and analyzing this small sample of obscured quasars are currently being used to understand the entire population of obscured quasars with both WISE and SDSS photometry. We can use our results from the photometric redshift estimation for these objects to aid future WISE-selected AGN clustering studies. The most heavily obscured quasars in our sample (such as J1354-0003, with a strong AGN component to the SED that only becomes apparent at the longest mid-IR wavelengths) may also be excellent targets for observations in the hard X-ray portion of the spectrum with The Nuclear Spectroscopic Telescope Array (\emph{NuSTAR}) High-energy X-Ray telescope \citep{harrison2013}. We are also using SALT to probe portions of the WISE W1$-$W2 vs. W2$-$W3 color-color diagram that are not spanned by our current sample. Together, our results along with future analysis help demonstrate the power of mid-IR WISE selection for recovering large samples of obscured quasars.

\acknowledgments 

We would like to thank the anonymous referee for their constructive comments which improved the final paper. KNH, RCH, ADM, and MAD were partially supported by NASA through ADAP award NNX12AE38G and by the National Science Foundation through grant numbers 1211096 and 1211112.

Some of the observations reported in this paper were obtained with the Southern African Large Telescope (SALT). This publication also makes use of data products from the Wide-field Infrared Survey Explorer, which is a joint project of the University of California, Los Angeles, and the Jet Propulsion Laboratory/California Institute of Technology, funded by the National Aeronautics and Space Administration. Funding for SDSS-III has been provided by the Alfred P. Sloan Foundation, the Participating Institutions, the National Science Foundation, and the U.S. Department of Energy Office of Science. The SDSS-III web site is http://www.sdss3.org/.

SDSS-III is managed by the Astrophysical Research Consortium for the Participating Institutions of the SDSS-III Collaboration including the University of Arizona, the Brazilian Participation Group, Brookhaven National Laboratory, Carnegie Mellon University, University of Florida, the French Participation Group, the German Participation Group, Harvard University, the Instituto de Astrofisica de Canarias, the Michigan State/Notre Dame/JINA Participation Group, Johns Hopkins University, Lawrence Berkeley National Laboratory, Max Planck Institute for Astrophysics, Max Planck Institute for Extraterrestrial Physics, New Mexico State University, New York University, Ohio State University, Pennsylvania State University, University of Portsmouth, Princeton University, the Spanish Participation Group, University of Tokyo, University of Utah, Vanderbilt University, University of Virginia, University of Washington, and Yale University.

\bibliographystyle{apj}


\end{document}